\documentclass[12pt]{article} 
\usepackage{graphicx} 
\usepackage{amssymb} 
\usepackage{amsmath}

\newcommand{\sechead}[1]{\par\vspace{0.5in}{\bf #1} \vspace{0.25in}}

\begin{document} 
\begin{titlepage}
	
	\vskip 2cm 
	\begin{center}
		\large{{\bf D3 branes in a Melvin universe:\\
		a new realm for gravitational holography}} 
	\end{center}
	
	\vskip 2cm 
	\begin{center}
		Julian Freed-Brown\footnote{\texttt{jfreedbrown@hmc.edu}} ,\ \ Vedika Khemani\footnote{\texttt{vkhemani@hmc.edu}} , and\ \ Vatche Sahakian\footnote{\texttt{sahakian@theory.caltech.edu}} \\
		\vskip 24pt {\sl \footnotemark[1]${}^,$\footnotemark[2]${}^,$\footnotemark[3] Harvey Mudd College, Claremont, CA 91711 USA} \\
		{\sl \footnotemark[3] California Institute of Technology, Pasadena, CA 91125, USA} 
	\end{center}
	
	\vskip 2cm 
	\begin{abstract}
		The decoupling limit of a certain configuration of D3 branes in a Melvin universe defines a sector of string theory known as Puff Field Theory (PFT) - a theory with non-local dynamics but without gravity. In this work, we present a systematic analysis of the non-local states of strongly coupled PFT using gravitational holography. And we are led to a remarkable new holographic dictionary. We show that the theory admits states that may be viewed as brane protrusions from the D3 brane worldvolume. The footprint of a protrusion has finite size - the scale of non-locality in the PFT - and corresponds to an operator insertion in the PFT. We compute correlators of these states, and we demonstrate that only part of the holographic bulk is explored by this computation. We then show that the remaining space holographically encodes the dynamics of the D3 brane tentacles. The two sectors are coupled: in this holographic description, this is realized via quantum entanglement across a holographic screen - a throat in the geometry - that splits the bulk into the two regions in question. We then propose a description of PFT through a direct product of two Fock spaces - akin to other non-local settings that employ quantum group structures.
	\end{abstract}
\end{titlepage}

\newpage \setcounter{page}{1}

\section{Introduction and results}

Non-local dynamics is the hallmark of string theory - the key ingredient for regularizing quantum gravity and providing for a consistent and unified UV completion of particle physics. In recent years, there have been several interesting attempts at zeroing onto this key attribute of the theory in simplified and computationally more accessible regimes. This involves a process of scaling out gravitational complications while retaining the non-local aspects of the parent theory. These attempts have come in several flavors: Non-Commutative Super Yang-Mills (NCSYM)~\cite{Seiberg:1999vs}-\cite{Maldacena:1999mh}, Non-Commutative Open String (NSOS) theory~\cite{Seiberg:2000ms}-\cite{Sahakian:2001xz}, Open Membrane (OM) theory~\cite{Gopakumar:2000ep}, Dipole Field (DF) theory~\cite{Bergman:2000cw}-\cite{Alishahiha:2002ex}, and, most recently, Puff Field theory (PFT) was proposed by Ganor~\cite{Ganor:2006ub,Ganor:2007qh,Haque:2008vm}. In all these instances, a decoupling limit is employed to scale out gravity and a flux or geometrical twist is used to tune the scale of the residual non-locality. PFT however stands out in that it can realize non-locality in an $SO(3)$ invariant manner, lending itself to cosmological applications.

To this date, relatively little is understood about PFT. In a recent work~\cite{Minton:2007fd}, a particular realization of a PFT was explored at strong coupling - through the holographic dual description - leading to a myriad of perplexing observations. The system seems to admit a very unusual holographic setup, one that necessitates going beyond the standard dictionary employed, for example, in the AdS/CFT context~\cite{Maldacena:1997re,Witten:1998qj,Gubser:1998bc}. \cite{Minton:2007fd} focused on cosmological implications of the results. In this work, we focus on the PFT itself, and unravel a surprisingly rich and beautiful holographic dictionary.

\sechead{On Puff Field Theory}

PFT is the worldvolume theory of D-branes in a Melvin universe~\cite{Ganor:2006ub}\footnote{For other realizations of branes in Melvin backgrounds, see~\cite{Cai:2006tda}-\cite{Dhokarh:2008ki}}. The setup can be realized in numerous different flavors: ones involving D-branes of different dimensionalities and varying amounts of supersymmetry (SUSY). The case we focus on in this paper is $3+1$ dimensional PFT with $\mathcal{N}=2$ SUSY and $U(1)\times U(2)$ R-symmetry, a theory of D3 branes in a Melvin background. The scale of non-locality arises from a single geometrical twist parameter. The theory flows in the IR to $\mathcal{N}=4$ local SYM.

An unusual aspect of this theory is that its spectrum includes non-local states - 3-ball bubbles of size set by the scale of non-locality in the theory. Yet the origin of such states is rather mysterious since there are no open D3 branes in the parent theory. And a Lagrangian description of this PFT is still missing.

\begin{figure}
	\begin{center}
		\includegraphics[width=7in]{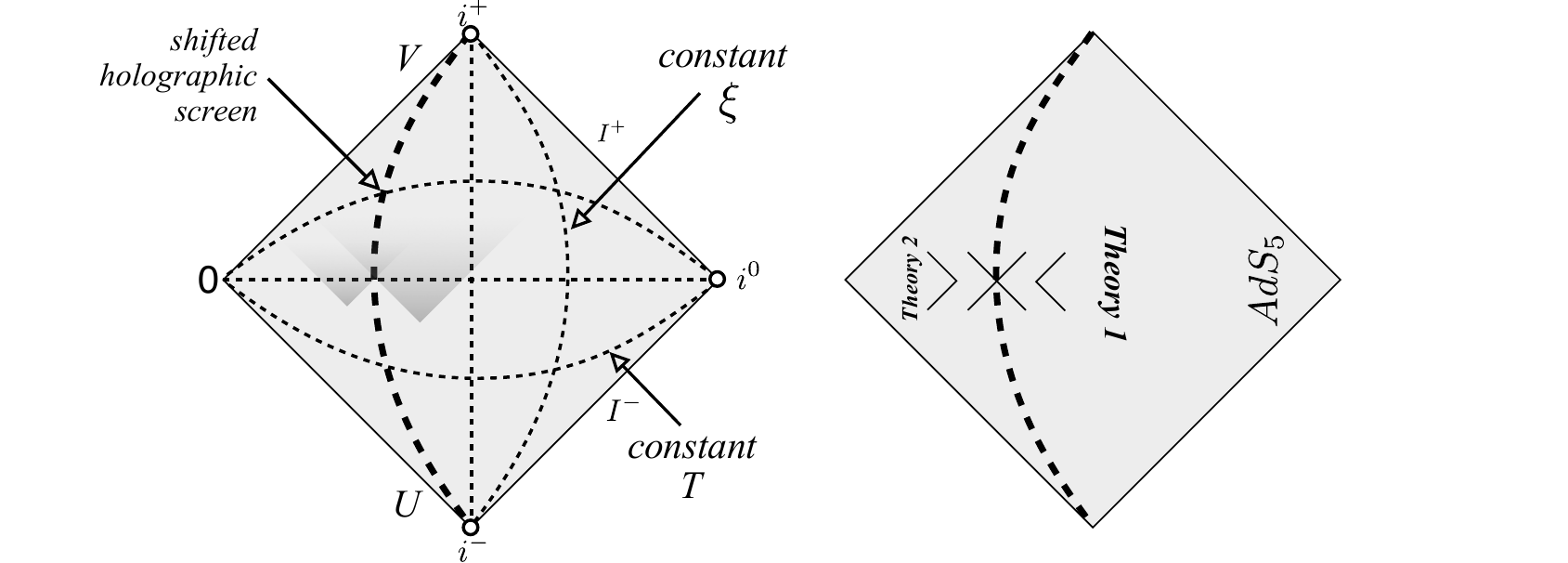} 
	\end{center}
	\caption{\em Penrose-Cartan diagram for the holographically dual geometry to our Puff Field Theory. Non-local operators in the PFT need to be inserted into the bulk space, at a new holographic screen. Bousso's criterion for holography suggests that the two sides of this new screen are holographically encoded onto their common boundary. The two shaded light-cones demonstrate the causal connectedness of the two sides of the holographic screen. On the right picture, solid wedges denote Bousso's light-sheets. The figure assumes that $\psi$ (that appears later on in the main text) is different from $\pi/2$ to avoid a singularity at $\xi=0$.} \label{fig:penrosecartan} 
\end{figure}

In~\cite{Ganor:2007qh}, the holographically dual geometry to this PFT was developed, making the strong coupling regime accessible through supergravity computations. The Penrose-Cartan diagram for the bulk spacetime is shown in Figure~\ref{fig:penrosecartan}.
We show on this diagram the PFT time coordinate $T$, and the holographic direction $\xi$. A standard UV-IR relation\cite{Peet:1998wn} matches high energy with small $\xi$, and low energy with large $\xi$
\begin{equation}
	\mu \sim \frac{1}{\xi}
\end{equation}
where $\mu$ is energy scale in the PFT. Local operators of the PFT are inserted in the UV, near $\xi=0$, as usual. And all seems holographically normal this far. The novelty arises when one considers operators with certain R-charges that are expected to have non-local features. They are the mysterious D3 bubbles alluded to earlier. In this work, we present a detailed exploration of such states, and find a remarkable holographic mechanism at work.

\newpage
\sechead{Summary of the results}

R-charge in the PFT is angular momentum in the holographic bulk. In this case, the bulk spacetime carries angular momentum as well, related to the geometrical twist that underlies the non-local character of the PFT. Measuring the R-charge of an operator insertion becomes a  delicate matter due to the effects of frame dragging. We show that, whenever the operator is to carry R-charge of a type expected to lead to non-local effects, the insertions would be placed {\em not} at the boundary $\xi=0$, but at $\xi=\xi_0>0$ inside the bulk - at a shifted holographic screen! To ascertain the reliability of this conclusion, we employ the covariant holographic criterion of Bousso~\cite{Bousso:1999xy}: we look at the rate of convergence of null geodesics projected in the holographic direction $\xi$. We find that this rate vanishes at $\xi=\xi_0$. Bousso's light-sheets are arranged such that the region $\xi>\xi_0$ is expected to be holographically encoded at $\xi=\xi_0$; while the region $\xi<\xi_0$ is also to be encoded on the {\em same surface} $\xi=\xi_0$. 

\begin{figure}
	\begin{center}
		\includegraphics[width=5in]{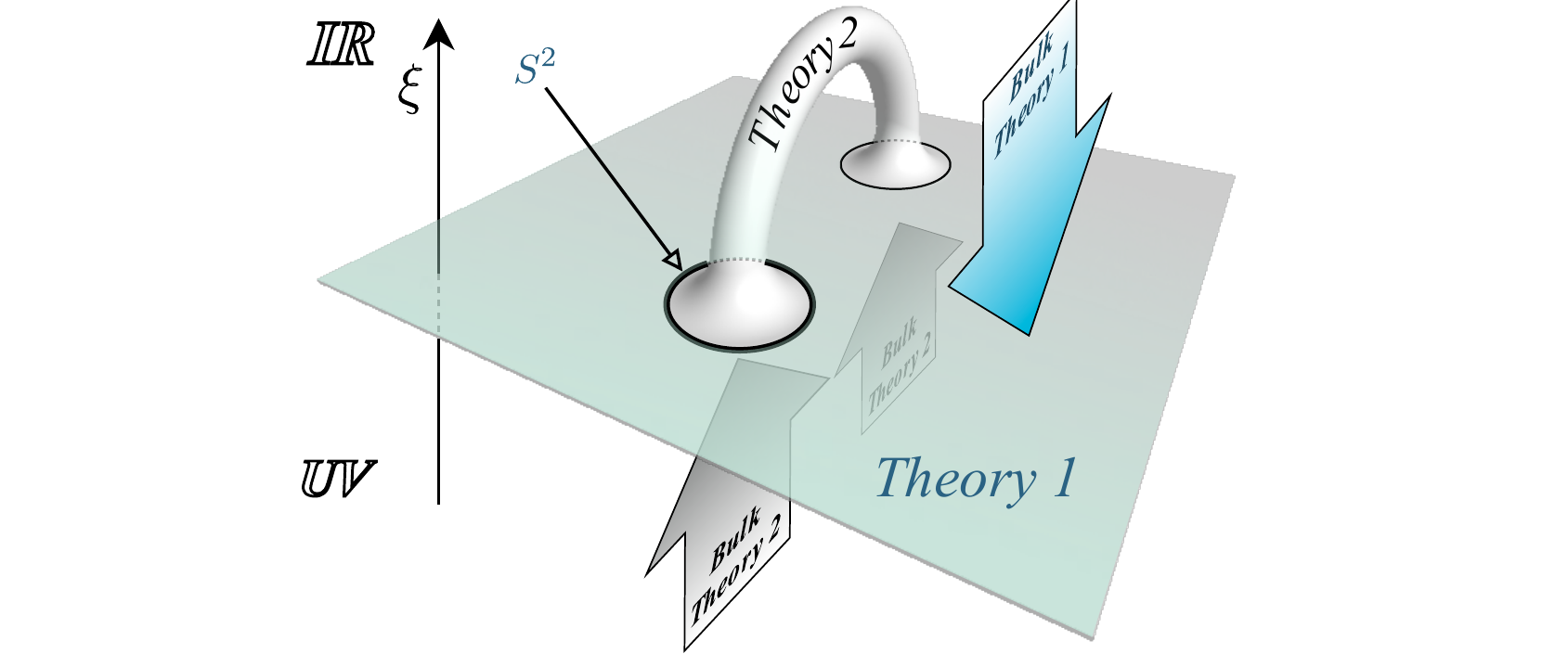} 
	\end{center}
	\caption{\em A cartoon for the insertion of two non-local operators in the PFT. Two bubbles are deleted from the worldvolume of the D3 branes, and joined by a D3 brane tentacle.} \label{fig:loop} 
\end{figure}

We then use geodesics to estimate the correlation functions between such non-local operator insertions. We find that the correlators involve a minimum distance scale $\Delta x_m$ 
\begin{equation}
	\langle \mathcal{O}(x_1) \mathcal{O}(x_2) \rangle \sim \left\{ 
	\begin{array}{ll}
		\left( \Delta x \right)^{-4\,h_+}  & \Delta x \gg \Delta x_m \\
		e^{-2\,h_+\, \left( \frac{\Delta x}{\Delta x_m}\right)} & \Delta x \,\widetilde{<}\, \Delta x_m 
	\end{array}
	\right. \ 
\end{equation}
for a bulk probe corresponding to an operator of weight $h_+$ in the IR of the theory. We show that $\Delta x_m$ corresponds to $\xi\sim \xi_0$ through the UV-IR relation, and is also the scale of non-locality in theory - proportional to the geometrical twist of the PFT. We are then led to the holographic dictionary depicted in Figure~\ref{fig:loop}.
In this Figure, $\xi$ is the vertical direction, and we cartoon the D3 brane worldvolume as a 2D surface. The computation of a two-point correlator of non-local states in the PFT involves a deformation of the worldvolume such that a D3 brane tentacle is projected into the bulk. The operator insertions may be viewed as two 3-balls excised from the worldvolume and joined by the tentacle. The holographic region $\xi>\xi_0$ describes the effective theory of interactions of the non-local footprints and may be viewed as projected onto the D3 worldvolume minus the two 3-bubbles. We call this in the Figure `Theory 1'.

The region $\xi<\xi_0$ on the other hand must describe the degrees of freedom of the tentacle! As seen from the Figure, it employs an {\em IR cutoff} related to the size of the 3-bubbles. We compute correlators along the tentacle through geodesics folding into the $\xi<\xi_0$ region. We demonstrate that higher energy in Theory 2 corresponds to larger separations between the operator insertions - a direct realization of UV-IR mixing: higher energy corresponds to a longer protrusion. The D3 tentacle may be viewed as the result of an open string pulling on its two endpoints that are attached to the D3 brane worldvolume. In general, we may consider `Theory 2' as the effective theory of such protrusions from the D3 brane worldvolume - states that will necessarily leave a non-local footprint in Theory 1.

In reality, Theory 1 and Theory 2 are coupled, as can be seen from the causal structure of the Penrose-Cartan diagram. The holographic screen at $\xi=\xi_0$ that separates the two theories in the bulk is where the convergence rate of null geodesics vanishes: it is a minimal volume neck in the geometry. From other unrelated work~\cite{Ryu:2006bv}-\cite{VanRaamsdonk:2009ar}, we know that such boundaries imply quantum entanglement between the two sides: when we refer to Theory 1, we are implicitly including effects of entanglement with the other side of the holographic screen, Theory 2. In the dual language, the effective theory of interaction of D3 bubbles encodes in it - through quantum entanglement - information about the dynamics of the internal degrees of freedom of the D3 tentacle, {\em i.e.} modes from the 3-spheres excised from the worldvolume. The full PFT theory must be the combination of the two theories. We speculate on the nature of this full theory as one involving a Fock space with direct product structure - dynamically generated mixed states of the two theories. We propose a specific mechanism for this phenomenon: a mixing inspired by a redshift effect in the dual bulk. The PFT's full Hamiltonian gets diagonalized only by certain combinations of states from each sector - akin to a similar mechanism employed in certain non-local theories based on quantum groups~\cite{Arzano:2007nx,Arzano:2008yc}.

We build up this narrative gradually, by starting with no preconceptions and employing several independent computations. Section 2 collects brief self-contained background material on the PFT. Section 3 sets up the basic tool we use to probe the bulk: geodesics with various angular momenta. Section 4 shows the relation between R-charge and the shifting of the holographic screen. Section 5 tests the waters with computations of 2-point correlators for operators expected to be local in character, and confirms these expectations. Section 6 explores 2-point correlators with non-local operators and collects the raw results. Section 7 demonstrates briefly a more complicated case involving correlators of different R-charges that requires numerical treatment. Section 8 presents the main analysis of the paper: it collects all the results in one place and systematically develops the conclusion summarized in this Introduction. Section 9 concludes the discussion and presents future directions for exploration. Finally, Appendix A collects a few short conventions about R-charges used in the main text.

\section{Background on Puff field theory}

\subsection{Definition of a PFT} \label{sub:definition}

Puff Field Theory is the worldvolume theory of D3 branes in a Melvin universe. We start in M-theory with Kaluza-Klein waves in the $z$ direction of the background metric~\cite{Ganor:2007qh}
\begin{equation}
	\label{eq:backgroundmetric} ds_{11}^2=-dt^2 +\sum_{i=1}^3 dx_i^2+\sum_{i=1}^3 \left( dr_i^2+r_i^2 \left( d\phi_i+\beta_i dz^2\ . \right)^2 \right)+dz^2\ ; 
\end{equation}
The twist in the geometry is parameterized by the three arbitrary constants labeled as $\beta_i$'s. To get to D3 branes, one reduces along the $z$ directions, T-dualizes along $x_1$, $x_2$, $x_3$, and takes the usual decoupling limit. The $\beta_i$'s tune the extent of non-locality in certain operators of the resulting worldvolume theory.

It is convenient to write the twist in the geometry through an $SO(6)$ matrix acting on the coordinates $r_i,\phi_i$ for $i=1,2,3$. Writing these as six cartesian coordinates $y_1,\cdots,y_6$, the twist matrix is (see Appendix A for more details)
\begin{equation}
	\label{eq:twist} \eta = \left( 
	\begin{array}{cccccc}
		0 & \beta_1 & 0 & 0 & 0 & 0 \\
		-\beta_1 & 0 & 0 & 0 & 0 & 0 \\
		0 & 0 & 0 & \beta_2 & 0 & 0 \\
		0 & 0 & -\beta_2 & 0 & 0 & 0 \\
		0 & 0 & 0 & 0 & 0 & \beta_3 \\
		0 & 0 & 0 & 0 & -\beta_3 & 0 
	\end{array}
	\right)\ .
\end{equation}
We focus on the special case where $\beta_1=\beta_2=\beta$ and $\beta_3=0$, yielding $\mathcal{N}=2$ supersymmetry in the corresponding PFT. We will come back to more general cases at the end of the paper.

Our PFT is a $3+1$ dimensional theory parameterized by: the number of D3 branes $N$, a large $N$ coupling constant $G\equiv 4\pi g_s N$ with $g_s$ being the string coupling in IIB theory, and a scale of non-locality $\Delta^3\equiv \beta {\alpha'}^2$. At low energies, this PFT is expected to flow to $\mathcal{N}=4$ Super Yang-Mills (SYM) with $SO(6)$ R-symmetry. The full theory however has the reduced R-symmetry $U(1)\times U(2)$ due to the twist, as is evident from~(\ref{eq:twist}). We will label the generators by ${\bf Q}_0$ for the $U(1)$, and ${\bf Q}_i$ with $i=1,\cdots,3$, and ${\bf Q}_4$ for the $U(2)$
\begin{equation}
	\label{eq:qcommutator} \left[ {\bf Q}_0, {\bf Q}_i \right]=0\ \ \ ,\ \ \ \left[ {\bf Q}_4, {\bf Q}_i \right]=0\ \ \ ,\ \ \ \left[ {\bf Q}_i, {\bf Q}_j \right]=\varepsilon_{ijk} {\bf Q}_k\ . 
\end{equation}
In~\cite{Ganor:2006ub}, it was argued that R-charged operators are expected to fuzz up in all three space directions to a volume proportional to $\Delta^3$, hence seeding non-local dynamics in the theory. From duality considerations, one expects that these states may be viewed as `D3 brane bubbles'. The resulting non-locality will preserve full $SO(3)$ symmetry in the $3+1$ dimensional theory, making the setup particularly attractive for cosmological applications. The `volume' of a state with charge matrix $J$ is defined by~\cite{Ganor:2006ub}
\begin{equation}
	V[J]\equiv \frac{1}{2} (2 \pi)^3 \mbox{Tr}[J\,\eta] {\alpha'}^2\ . 
\end{equation}
For our case, this volume expression vanishes except when $Q_4$ is activated 
\begin{equation}
	\label{eq:volumeq4} V[q {\bf Q}_4]=8 \pi^3 q {\Delta^3} 
\end{equation}
for some arbitrary charge $q$ in the ${\bf Q}_4$ direction. We will see the significance of this later on in the geodesic probe analysis.

A full Lagrangian description of our PFT is still missing; but we will only be concerned with the strong coupling regime of the theory. And the latter is accessible through holography by studying supergravity excitations in a particular curved background. We  focus on the strong coupling regime of the PFT for the rest of the paper.

\subsection{Holographic dual} \label{sub:holographic_dual}

Following the construction described in the previous section, the holographic dual geometry to the PFT of interest was derived in~\cite{Ganor:2007qh}. The IIB string metric is given by 
\begin{eqnarray}
	\label{eq:thebigmetric} ds_{str}^2&=& \alpha' K^{1/2} \left[ -H^{-1} dt^2 +dV^2+ V^2 d\psi^2 +\frac{V^2 \cos^2\psi}{4} \left( d\theta^2+\sin^2 \theta\, d\varphi^2\right) +V^2 \sin^2 \psi\, d\chi^2 \right] \nonumber \\
	&+& \alpha' K^{-1/2} \left[ \sum_i dx_i^2 + H\, V^2 \cos^2\psi\, \left( d\phi -\frac{1}{2} \left( 1-\cos\theta \right)d\varphi + \Delta^3 H^{-1} dt \right)^2 \right]
\end{eqnarray}
incorporating the back-reaction of the D3 branes.
The worldvolume of the D3 branes spans the coordinates $x_i$ with $i=1,2,3$ and the time direction $t$; the holographic direction associated with energy scale in the dual PFT is denoted by $V$; and the remaining five directions are compact and denoted by the angles $\{\phi$, $\varphi$, $\psi$, $\theta, \chi\}$\footnote{To relate these angles to the setup in~\cite{Ganor:2007qh} and the previous section's discussion, see Appendix A.}. The $\theta$, $\phi$, and $\varphi$ parameterize an $S^3$ which may be viewed as a Hopf fibration with $\phi$ labeling the fiber direction and the base $S^2$ described by $\{\theta,\varphi\}$. The bounds on the angular coordinates are: 
\begin{equation}
	0\leq\psi\leq\pi/2\ \ ,\ \ 0\leq\theta\leq\pi\ \ ,\ \ 0\leq\varphi\leq 2\,\pi\ \ ,\ \ 0\leq\chi\leq 2\,\pi\ \ ,\ \ 0\leq\phi\leq 2\,\pi \ .
\end{equation}
The two factors $H$ and $K$ appearing in the metric are given by 
\begin{equation}
	H=\frac{4\pi g_s N}{V^4}\ \ \ \mbox{and}\ \ \ K=\frac{4\pi g_s N}{V^4}+\Delta^6 V^2 \cos^2\psi \ , 
\end{equation}
where $g_s$ is the IIB string coupling, $N$ is the number of D3 branes, and $\Delta$ is the scale of non-locality (and has length dimension $1$). The dilaton profile is constant 
\begin{equation}
	e^\Phi = g_s 
\end{equation}
but there is a non-trivial 5-form RR flux that supports the geometry. The latter is of no relevance to our discussion as we will concentrate on supergravity probes with zero RR charge.

For convenience, and to exhibit the relevant physical scales in the problem, apply a coordinate change 
\begin{equation}
	z=\frac{1}{V} \ , 
\end{equation}
and rescale the coordinates to dimensionless variables 
\begin{equation}
	\label{eq:rescalings} \xi\equiv \frac{G^{1/6}}{\Delta} z\ \ \ ,\ \ \ X_i\equiv \frac{x_i}{G^{1/3} \Delta}\ \ \ ,\ \ \ T \equiv \frac{t}{G^{1/3} \Delta} \ , 
\end{equation}
where 
\begin{equation}
	G\equiv 4\pi g_s N \ . 
\end{equation}
We will later need to take $N\gg 1$, with $G$ becoming the effective coupling in the PFT. In the new coordinates, the metric is 
\begin{eqnarray}
	\label{eq:puffmetric} ds_{str}^2 &=&\alpha'\, \sqrt{G}\, \frac{\sqrt{\xi^6+\cos^2\psi}}{\xi^5}\left[ -dT^2+d\xi^2+\xi^2d\psi^2+\xi^2 \frac{\cos^2\psi}{4} \left( d\theta^2+\sin^2\theta d\varphi^2 \right)+\xi^2\sin^2\psi\, d\chi^2 \right] \nonumber \\
	&+& \alpha' \sqrt{G} \frac{\xi}{\sqrt{\xi^6+\cos^2\psi}} \left[ dX_i^2+\xi^2 \cos^2\psi \left( d\phi -\frac{1}{2} (1-\cos\theta) d\varphi+\frac{dT}{\xi^4} \right)^2 \right] \ . 
\end{eqnarray}
For large $\xi$, the geometry asymptotes to $AdS_5\times S^5$ with cosmological constant $\Lambda={4}/{\alpha' \sqrt{G}}$; hence the theory flows to $\mathcal{N}=4$ SYM with central charge $N^2/4$. As $\psi\rightarrow \pi/2$, we encounter a singularity at $\xi=0$, which we will come back to later.

The decoupling limit for this setup involves taking $\alpha'\rightarrow 0$ while holding $G$, $\xi$, $T$, $X_i$ fixed.

\sechead{Holographic screen}

Gravitational holography can be unravelled most easily using Bousso's covariant criterion. One looks at the rate of convergence of null geodesics projected transverse to the worldvolume of the branes; negative or zero convergence defines a light-sheet and a corresponding region of bulk space that is candidate for holographic encoding. In our case, the problem is somewhat more subtle as the spacetime carries angular momentum and drags geodesics with it. If we consider null geodesics with zero angular momentum, one can show that (using the Killing vectors of Section~\ref{sec:geodesics}) the tangent $k^a$ to the geodesics should be 
\begin{equation}
	k^a=\left( \frac{|{E}| \sqrt{\xi ^6+\cos ^2\psi}}{\xi }+\frac{{E} \cos ^2\psi}{\xi \sqrt{\xi ^6+\cos ^2\psi}} \right) 
	\partial_T\pm\frac{ \xi ^5 |{E}|}{\sqrt{\xi ^6+\cos ^2\psi}} 
	\partial_\xi+\frac{{E}\, \xi }{\sqrt{\xi ^6+\cos ^2\psi}}
	\partial_\phi\ , 
\end{equation}
where $E$ is energy and the upper sign in $\pm$ refers to geodesics projected towards the $AdS_5$ region, at larger values of $\xi$ ({\em i.e.} `ingoing' geodesics); and the lower sign refers to geodesics pointing towards $\xi=0$ ({\em i.e.} `outgoing' geodesics). The sign of $E$ determines whether the projection is future ($E>0$) directed, or past ($E<0$) directed. The geodesics have to spin in the $\phi$ direction to make up for the dragging effect due to the non-zero angular momentum of the background spacetime. Note also the dependence of this expression on the $\psi$ angle, which we take at first as constant but otherwise tunable between $0$ and $\pi/2$. 
\begin{figure}
	\begin{center}
		\includegraphics[width=5.5in]{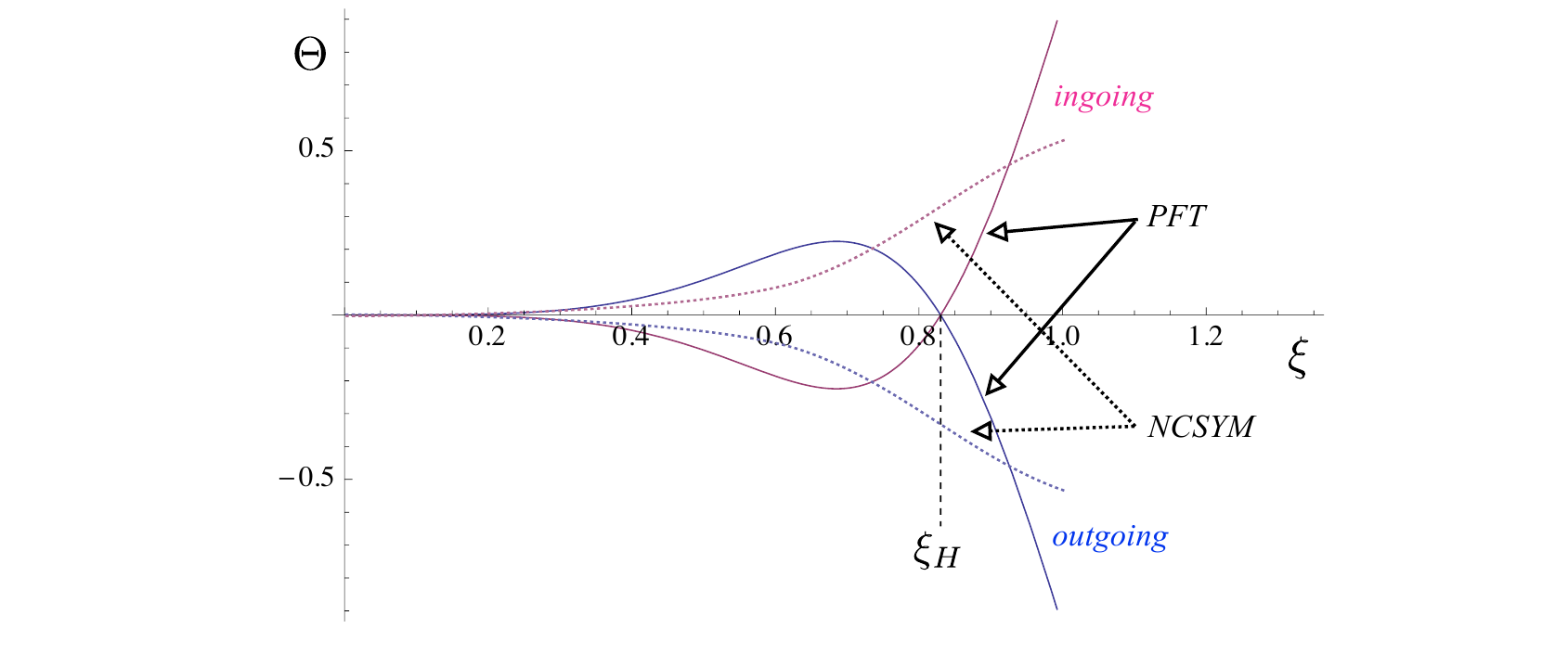} 
	\end{center}
	\caption{\em The holographic screen determined as the point where the convergence rate of null geodesics vanishes, $\Theta=0$. `Outgoing' refers to null geodesics projected towards larger values of $\xi$; and `ingoing' refers to null geodesics projected towards $\xi=0$. To demonstrate the unusual behavior of the PFT, we also show the corresponding quantity (dashed curves) for the case of $3+1$ dimensional Non-Commutative Super Yand-Mills.} \label{fig:holographicscreen} 
\end{figure}
The convergence rate is then\footnote{ We could also consider null geodesics with non-zero angular momentum; for example, labeling one of the angular momenta as $\mathcal{L}_3$, one gets 
\begin{equation}
	\Theta = \pm \frac{3 \xi ^3 \left(\cos ^2\psi -2 \xi ^6\right) \sqrt{{E}^2 \xi ^2-4 \mathcal{L}_3^2 \sec ^2\psi}}{2 \left(\xi ^6+\cos ^2\psi\right)^{3/2}} \ .
\end{equation}
The sign of $\Theta$ is still determined by the $(\cos ^2\psi -2\, \xi ^6)$ factor. And we see the inclusion of the angular momentum does not change the conclusion about holography obtained from the zero angular momentum case. } 
\begin{equation}
	\label{eq:thetheta} \Theta \equiv \nabla_i k^i= \pm |E|\, \frac{3\, \xi^4 \left(\cos^2\psi-2\, \xi^6\right)}{2 \left(\xi^6+\cos^2{\psi}\right)^{3/2}} 
\end{equation}
where $i$ sums over the $x_1$, $x_2$, and $x_3$ directions. This leads to the identification of a critical point 
\begin{equation}
	\label{eq:holographicscreen} \xi=\xi_H\equiv \left( \frac{\cos^2\psi}{2}  \right)^{1/6}
\end{equation}
where $\Theta=0$ as shown in Figure~\ref{fig:holographicscreen}. The region of the bulk where $\xi>\xi_H$ is then candidate for holographic encoding at the surface $\xi=\xi_H$; and the region where $\xi<\xi_H$ is also a candidate for holographic encoding {\em on the same surface} $\xi=\xi_H$. We also see that, as we move the plane of the null geodesics from $\psi=0$ to $\psi=\pi/2$, the screen moves towards $\xi_H=0$. We will later demonstrate a relation between this shifting of the holographic screen and the R-charge of the operator inserted in the dual PFT.

It is useful to also note that the vanishing of $\Theta$ is a statement about the rate of change of the transverse volume to the geodesics: the region where $\Theta=0$ is a manifold of minimal volume in the $x_1$-$x_2$-$x_3$ subspace. Around this critical point, this volume increases as we move towards small $\xi$ {\em and} large $\xi$. Manifolds of minimal volume play a special role in holography - as boundaries that bound bulk regions in correspondence with dual entangled states~\cite{Ryu:2006bv}-\cite{Nishioka:2009un}.

\sechead{Regime of validity}

We restrict our analysis of the PFT to a low energy supergravity regime in the bulk. Additionally, we employ the optical approximation: we will use geodesics as probes of dual correlation functions in the PFT. Hence, we need to make sure that the setup is consistent and reliable within this setting. The first condition arises from requiring that the string frame curvature scale is small compared to the string scale. We look at a curvature invariant such as~\cite{Ganor:2007qh}
\begin{eqnarray}
	\label{eq:curvature} && R_{\mu\nu\alpha\beta}R^{\mu\nu\alpha\beta}=\frac{\xi ^6}{{\alpha'}^2G \left(\xi ^6+\cos ^2\psi\right)^5}\times \nonumber \\
	&& \left[-576\, \xi ^6 \cos ^6\psi+\left(8 \left(408\, \xi ^6+7\right) \xi ^6+65\right) \cos ^4\psi+24 \left(-10\, \xi ^{12}+29\, \xi ^6+5\right) \xi ^6 \cos ^2\psi \right. \nonumber \\
	&& \left. +40 \left(2 \left(\xi^6-5\right) \xi ^6+3\right) \xi ^{12}\right] 
\end{eqnarray}
and require that $R_{\mu\nu\alpha\beta}R^{\mu\nu\alpha\beta}\gg {\alpha'}^{-2}$. Equation~(\ref{eq:curvature}) exhibits a singularity when both $\xi$ and $\cos\psi$ vanish. Since we will be interested in approaching the geometry from $\psi<\pi/2$, we consider the $\xi\rightarrow 0$ limit first, at fixed $\psi<\pi/2$ 
\begin{equation}
	R_{\mu\nu\alpha\beta}R^{\mu\nu\alpha\beta}=\frac{65\, \xi^6}{{\alpha'}^2 G \cos^6 \psi} \ .
\end{equation}
For small curvatures compared to the string scale, we then need 
\begin{equation}
	\label{eq:condition1} \xi\ll G^{1/6} \cos \psi\ . 
\end{equation}
We will later find that $\xi$ is bounded from above by numerical values independent of the parameters in the theory. Equation~(\ref{eq:condition1}) then implies a restriction on $\psi$ 
\begin{equation}
	\label{eq:conditionpsi} \cos\psi \gg G^{-1/6} \ .
\end{equation}
This hence leads to the generic strong coupling condition for holography 
\begin{equation}
	\label{eq:conditionG} G \gg 1\ . 
\end{equation}

We also need to make sure that the circle along the $\phi$ direction involved in the original twist is not too small compared to the string scale. This leads to the condition 
\begin{equation}
	\label{eq:tduality} \sqrt{G} \frac{\xi^3\cos^2\psi}{\sqrt{\xi^6+\cos^2\psi}} \gg 1 
\end{equation}
which adds a lower bound on $\xi$ 
\begin{equation}
	\label{eq:conditionxi} \xi \gg G^{-1/9} \ .
\end{equation}
Otherwise, we would need to consider the T-dual geometry, which is certainly a tractable issue but will be unnecessary for our purposes. Comparing the lower bounds on $\xi$ and $\cos \psi$ given by~(\ref{eq:conditionpsi}) and~(\ref{eq:conditionxi}) along the holographic screen $\xi_H^6=\cos^2\psi/2$, we note that the bound on $\cos \psi$ is more stringent for large $G$. Hence, the holographic screen gets cut off at $\cos^2 \psi \sim \xi^6 \sim G^{-1/3}\ll 1$. In~\cite{Ganor:2007qh}, the $x_1$, $x_2$, and $x_3$ directions were also compactified on a torus and the corresponding T-duality conditions were considered. In our case, we will not consider the PFT in a box, and hence will not have such additional conditions on our geometry.

\begin{figure}
	\begin{center}
		\includegraphics[width=8in]{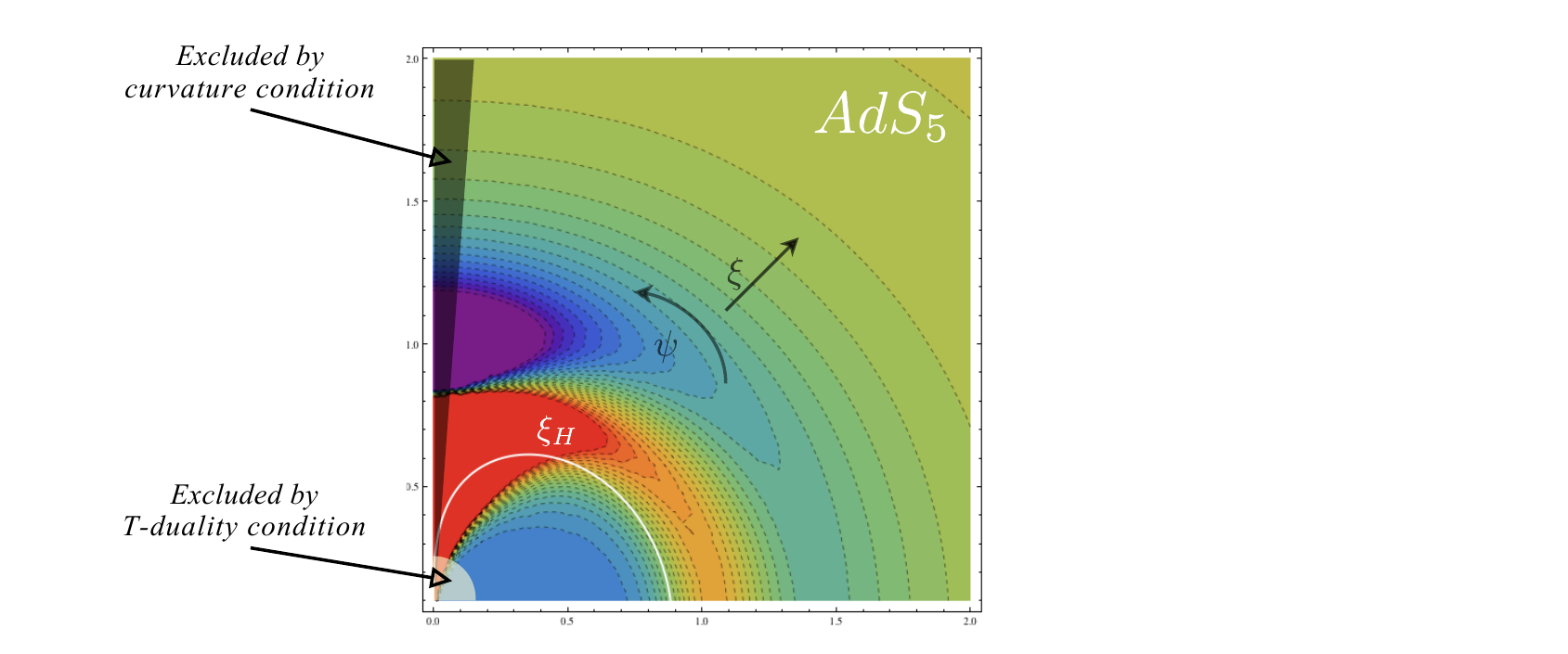} 
	\end{center}
	\caption{\em Contour plot of the curvature scale (equation~(\ref{eq:curvature})) in the $\xi$-$\psi$ directions. The layout is polar, with $0<\psi<\pi/2$ being the polar angle and $\xi$ being the radial direction. Blue (darker) areas correspond to negative values of $R_{\mu\nu\alpha\beta}R^{\mu\nu\alpha\beta}$, while red (lighter) areas to positive values. The conditions for small curvature and large $\phi$ circle are shown as well. The holographic screen $\xi_H$ is shown as a white line separating the space into two regions. The rate of convergence of null geodesics vanishes at $\xi_H$.} \label{fig:curvatureplot} 
\end{figure}

Finally, we also require weak string coupling 
\begin{equation}
	\label{eq:weakstringcoupling} g_s= \frac{G}{4\pi N}\ll 1 \ .
\end{equation}
Using~(\ref{eq:conditionG}), this implies that one needs $N\gg 1$. Hence, the bulk geometry is subject to the independent bounds given by~(\ref{eq:conditionpsi}),~(\ref{eq:conditionxi}), and~(\ref{eq:weakstringcoupling}). These conclusions are depicted in Figure~\ref{fig:curvatureplot}. 

As we will restrict our computations to using geodesic probes, the optical approximation is implicit throughout. It is valid when the wavelength of the bulk probe is much smaller than the local curvature of the space it is moving through. At the same time, we need the probe to be light enough not to back-react on the geometry. These conditions were analyzed in detail in~\cite{Minton:2007fd} and, for a probe of mass $m$, lead to 
\begin{equation}
	m \sqrt{\alpha' \sqrt{G}} \gg 1\ . 
\end{equation}
The scale $\sqrt{\alpha' \sqrt{G}}$ will be important to our upcoming discussion; we henceforth define 
\begin{equation}
	R^2 \equiv \alpha' \sqrt{G}\ . 
\end{equation}
and we write $m\, R\gg 1$. The cosmological constant of the geometry at large $\xi$ where the space asymptotes to $AdS_5$ is $4/R^2$. In the $AdS_{d+1}$/CFT duality, an operator of dimension $2\, h_+$ corresponds to a bulk probe of mass $m$ such that~\cite{Witten:1998qj,Balasubramanian:1998sn,Balasubramanian:1998de}
\begin{equation}
	h_+=\frac{d}{4}+\frac{\sqrt{d^2+4\, m^2 R^2}}{4} 
\end{equation}
Hence, with $m\, R\gg 1$, we would have 
\begin{equation}
	\label{eq:weight} h_+\simeq \frac{1}{2} m R
\end{equation}
at the IR fixed point.

\sechead{UV-IR relation and thermodynamics}

In~\cite{Ganor:2007qh}, the finite temperature realization of~(\ref{eq:puffmetric}) was also considered. As usual, it is given by insertions of horizon generating factors in $g_{TT}$ and $g_{\xi\xi}$, leading to a black hole with finite temperature 
\begin{equation}
	\mbox{Temp} = \frac{1}{\pi\,{G}^{1/3} \Delta} \frac{1}{\xi_h} \ ,
\end{equation}
where $\xi=\xi_h$ is the location of the horizon\footnote{ Note that the thermodynamics of the black hole must be computed in the Einstein frame metric $ds_{Ein}^2=g_s^{-1/2} ds_{str}^2$.
}. This helps us identify a UV-IR relation between energy scale $\mu$ in the PFT and extent in the bulk $\xi$ 
\begin{equation}
	\label{eq:uvir} \frac{\mu}{\mu_{nl}}\equiv \frac{1}{\xi} 
\end{equation}
where we also defined an energy scale of non-locality 
\begin{equation}
	\mu_{nl}\equiv \frac{1}{G^{1/3} \Delta}\ . 
\end{equation}
We will later see that $\mu_{nl}$ sets indeed the energy scale at which non-locality ensues. Notice that, at strong coupling $G$, the implication is that weak coupling non-locality scale $\Delta$ is enhanced at strong coupling by a power of $G$. 

The entropy of the black geometry is given by 
\begin{equation}
	S=N^2 (2\pi)^3 V_3 T^3 \ ,
\end{equation}
where $V_3$ is the volume of the $x_1$, $x_2$, and $x_3$ directions in a IR regularization scheme. This expression is interestingly identical to the finite temperature state of the $\mathcal{N}=4$ local SYM theory. It has to account for all degrees of freedom of the PFT with given asymptotic charges.

\newpage
\sechead{Penrose-Cartan diagram}

The Penrose-Cartan diagram can be presented most easily in the $T$-$\xi$ plane by first introducing a new coordinate $\tilde{\xi}$ such that 
\begin{equation}
	\frac{d\tilde{\xi}}{d\xi}=\sqrt{1+\frac{\cos^2\psi}{\xi^6}}\ . 
\end{equation}
We would then define the usual `null directions' as in 
\begin{equation}
	u= T-\tilde{\xi}\ \ \ ,\ \ \ v= T+\tilde{\xi}\ ; 
\end{equation}
and then compactify by 
\begin{equation}
	\tan U =e^{-u}\ \ \ ,\ \ \ \tan V=e^v\ , 
\end{equation}
leading to the relevant part of the metric taking the form
\begin{equation}
	ds^2\rightarrow \frac{\alpha' \sqrt{G}\, \xi}{\sqrt{\cos^2\psi+\xi^6}} \sec^2 U\, \sec^2\, V e^{-2\,\tilde{\xi}} dU\, dV\ . 
\end{equation}
The corresponding diagram is shown in Figure~\ref{fig:penrosecartan}, along with the location of the holographic screen $\xi_H$ and Bousso's light-sheets~\cite{Bousso:1999xy} that determine the regions that are candidates for holographic projection. At every point in this diagram, there are three non-compact spatial directions $x_1$, $x_2$, and $x_3$ that make up the space directions of the dual PFT theory, and five angles. One of these angles, $\psi$, plays a special role in that it tunes the location of the holographic screen. Operator insertions at various values of $\psi$ will be seen to carry different amounts of R-charge due to the spin of the background space. The other four angles describe a compact manifold with $U(1)\times U(2)$ isometry.

\section{Geodesics} \label{sec:geodesics}

A useful observable for exploring the effects of non-locality in a theory is the equal-time 2-point correlator: the vacuum expectation value of two insertions of an operator separated by some distance in space but otherwise at the same instant in time. In a conformal theory, this measurable is entirely determined by the conformal weight $h_+$ of the operator 
\begin{equation}
	\left< \mathcal{O}(x_a)\mathcal{O}(x_b)\right> \sim \frac{1}{\Delta x^{4 h_+}} 
\end{equation}
due to the absence of any length scales in the theory (writing $\Delta x \equiv |x_a-x_b|$). If the theory involves a massive particle, or another natural scale such as one arising from a length scale of non-locality, we may encounter exponential factors of the form $e^{\pm\Delta x/\Delta x_0}$, where $\Delta x_0$ is determined by the mass scale of a particle, or the energy scale at which non-local effects become important (see for example~\cite{Rozali:2000np,Gross:2000ba}). 

To compute the equal-time correlator through the holographic dictionary, one can employ the supergravity approximation in the bulk - valid at low enough energies; and at another coarser level of approximation, one can use geodesics. These would connect the points of operator insertions in the boundary theory. For equal time correlators, these are spacelike geodesics that touch the boundary of the bulk spacetime at two places. The length of such a geodesic $l$ - a function of $\Delta x$ - gives us the leading tree level contribution to the correlator 
\begin{equation}
	\left< \mathcal{O}(x_a)\mathcal{O}(x_b)\right> \sim e^{-m\, l}\ . 
\end{equation}
Here, $m$ is the mass of a state in the bulk theory associated with the boundary operator in question. In an AdS/CFT setup, $m$ is related to the weight of the operator by~(\ref{eq:weight}). 

\begin{figure}
	\begin{center}
		\includegraphics[width=6in]{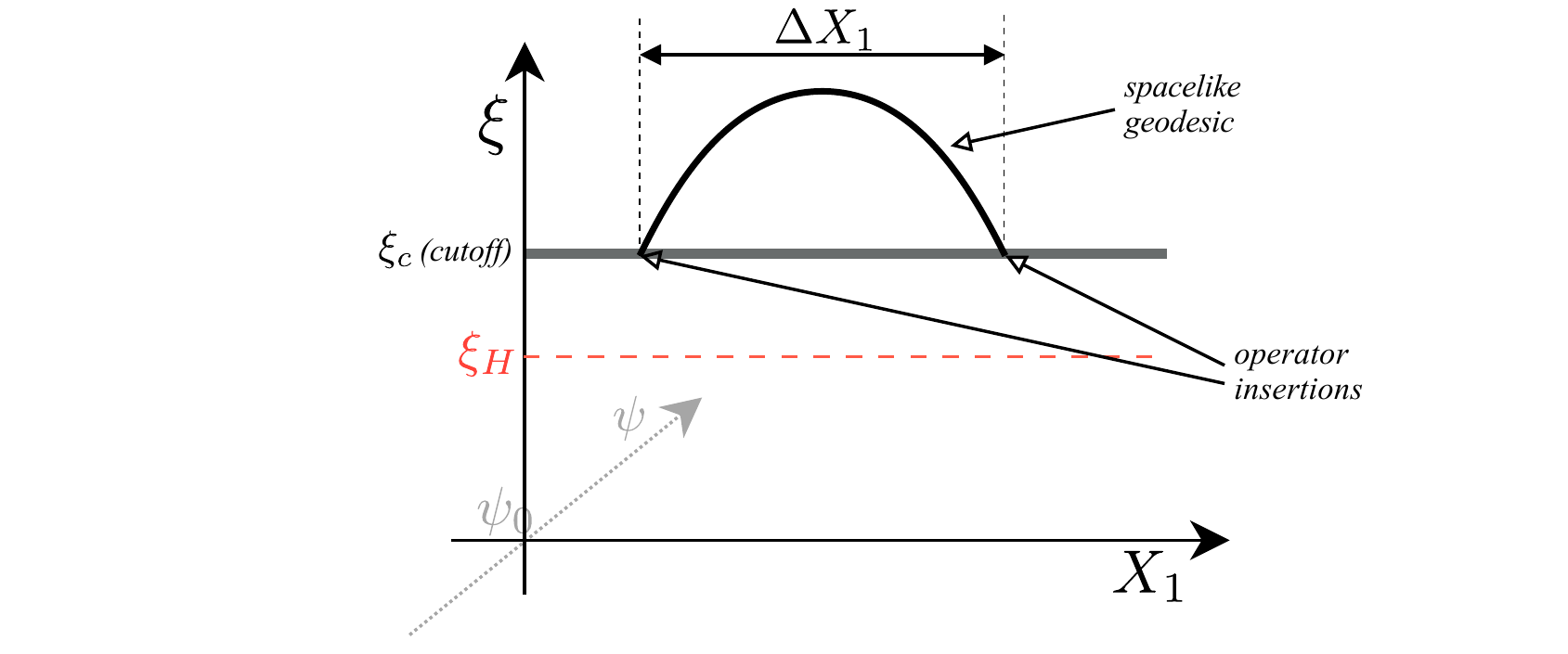} 
	\end{center}
	\caption{\em A spacelike geodesic connecting two points at a UV cutoff $\xi=\xi_c$ in the $\psi=\psi_0$ plane. The length of this geodesic is related to the vacuum expectation value of two operator insertions in the PFT.} \label{fig:holography} 
\end{figure}

The bulk points where the operators are inserted at also determine the UV cutoff employed in the computation of the correlator in the dual theory: typically, the UV-IR correspondence associates an energy in the boundary theory with extent along a `holographic' direction in the bulk transverse to the boundary. In our case, the UV-IR relation~(\ref{eq:uvir}) identifies the holographic direction as the $\xi$ coordinate. In Figure~\ref{fig:holography}, we show a cartoon of the setup: $\mu_c=\mu_{nl}/\xi_c$ is a UV-cutoff and hence determines the location $\xi_c$ of the holographic screen, and the IR of the theory lies at larger values of $\xi$. Two operators are inserted at this holographic screen and form the endpoints of a spacelike geodesic bending into the IR region. 

In our case, the situation is more interesting and complicated. The insertions are sensitive to the $\psi$ angular coordinate at which they lie: due to the spinning background geometry, insertions will carry angular momenta (and hence the corresponding operator will carry R-charge) of varied amounts depending on where they are inserted on the compact manifold transverse to $T$, $\xi$, and $X_1,X_2,X_3$. We depict this in the figure by adding a third dimension in and out of the page for $\psi$. Furthermore, the bulk spacetime has a natural holographic screen that shifts with $\psi$ from $\xi=0$ at $\psi=\pi/2$ to $\xi=2^{-1/6}$ for $\psi=0$. Bousso's criterion for holography implies the existence of two regions about the $\xi=\xi_H$ surface available for holographic projection. We will elaborate on these points in the next section. For now, let us proceed with the traditional tools of analysis of geodesics in curved spacetimes.

\newpage
\sechead{The Killing vectors}

It is a straightforward exercise to map out the Killing vectors of the metric~(\ref{eq:puffmetric}). Labeling these in correspondence with the notation used for the R-symmetry generators in~(\ref{eq:qcommutator}), we write 
\begin{equation}
	l_0=
	\partial_\chi \ ;
\end{equation}
\begin{equation}
	l_1=\cot {\theta} \sin {\varphi}\,
	\partial_\varphi-\frac{1}{2} \tan (\theta/2) \sin {\varphi}\, 
	\partial_\phi-\cos {\varphi}\,
	\partial_\theta \ ;
\end{equation}
\begin{equation}
	l_2=-\cot {\theta} \cos {\varphi}\,
	\partial_\varphi+\frac{1}{2} \tan (\theta/2) \cos {\varphi}\,
	\partial_\phi-\sin\varphi\,
	\partial_\theta \ ;
\end{equation}
\begin{equation}
	l_3=-\frac{1}{2}
	\partial_\phi-
	\partial_\varphi \ ;
\end{equation}
\begin{equation}
	l_4=
	\partial_\phi \ .
\end{equation}
The momentum Killing vector with respect to the rescaled $X_i$ coordinate is 
\begin{equation}
	p_i=
	\partial_{X_i} 
\end{equation}
where $i=1,2,3$. Given the Poincar\'{e} symmetry of the $x_1$-$x_2$-$x_3$ plane, we henceforth focus on one of the three directions, say $X\equiv X_1$, without loss of generality; and we write $p=
\partial_X$. For energy with respect to the rescaled $T$ coordinate, we write 
\begin{equation}
	e=
	\partial_T\ . 
\end{equation}

\sechead{Conservation laws}

To map out the spacelike geodesics of interest, we use the conservation equations 
\begin{equation}
	\mathcal{L}_i={l}_{i}^a v_a\ \ \ ,\ \ \ i=0\cdots 4 
\end{equation}
where $v_a$ is the tangent to the geodesic satisfying $v_a v^a=1$, and we adopt the ansatz $v^T=0$ with all other components non-vanishing. The $\mathcal{L}_i$'s are constants of motion with length dimension one. Similarly, we have 
\begin{equation}
	\mathcal{E}=e^a v_a\ \ \ \ , \ \ \ \mathcal{P}=p^a v_a 
\end{equation}
with $\mathcal{E}$ and $\mathcal{P}$ having length dimension one as well. We will use the shorthand $'\rightarrow d/d\tau$, where $\tau$ is the affine parameter along the geodesic. And we choose units so that $R=1$. After all these preparations, we obtain the following first order differential equations 
\begin{equation}
	\label{eq:chi} \chi'= \frac{{\mathcal{L}_0} \xi^3 \csc^2{\psi}}{\sqrt{\xi^6+\cos^2{\psi}}} \ ;
\end{equation}
\begin{equation}
	\theta'= -\frac{4\, \xi^3 \sec^2{\psi}\, ({\mathcal{L}_1} \cos {\varphi}+{\mathcal{L}_2} \sin {\varphi})}{\sqrt{\xi^6+\cos^2\psi}} \ ;
\end{equation}
\begin{equation}
	\phi '=\frac{\sin \left(\theta/2\right) \csc \theta \left(\xi ^6 (\mathcal{L}_4-2 \mathcal{L}_3) \sec \left(\theta/2\right) \sec ^2\psi+2 \mathcal{L}_4 \cos \left(\theta/2\right)\right)}{\xi ^3 \sqrt{\xi ^6+\cos ^2\psi}} \ ;
\end{equation}
\begin{equation}
	\varphi'= -\frac{4\, \xi^3 \sec^2{\psi}\, (\cot {\theta}\, ({\mathcal{L}_2} \cos {\varphi}-{\mathcal{L}_1} \sin {\varphi})+{\mathcal{L}_3})}{\sqrt{\xi^6+\cos^2\psi}} \ ;
\end{equation}
\begin{equation}
	{X'}= \frac{\mathcal{P} \sqrt{\xi^6+\cos^2{\psi}}}{\xi } \ .
\end{equation}
Finally, the normalization condition $v^a v_a=1$ leads to 
\begin{eqnarray}
	\label{eq:norm} && \xi ^6 \sec ^2\theta \left(\xi ^6 \sec ^2\left(\frac{\theta}{2}\right) \sec ^2\psi+1\right) \mathcal{L}_5^2+ \xi^{12} \sec ^4\left(\frac{\theta}{2}\right) \sec ^2\psi \mathcal{L}_3^2 \nonumber \\
	&& +2 \xi ^{12} \tan \left(\frac{\theta}{2}\right) \sec ^2\left(\frac{\theta}{2}\right) \sec \theta \sec ^2\psi \mathcal{L}_5 \mathcal{L}_3 \nonumber \\
	&& +4 \tan ^2\left(\frac{\theta}{2}\right) \left(-\cos ^2\psi (\mathcal{L}_1 \sin \varphi+2 \mathcal{L}_3 \cot \theta)+\mathcal{L}_2 \cos \varphi \cos ^2\psi-2 \mathcal{L}_3\, \xi ^6 \csc \theta \right) \mathcal{L}_5 \nonumber \\
	&& + \mathcal{L}_3^2 \sec ^4\left(\frac{\theta}{2}\right) \cos \theta \left(2\, \xi ^6 +\cos \theta \cos ^2 \psi\right) +\frac{\sqrt{\xi ^6+\cos ^2\psi}}{\xi ^3}\left(\psi'^2+\frac{\xi'^2}{\xi^2}\right)=1 
\end{eqnarray}
where we defined 
\begin{equation}
	\mathcal{L}_5\equiv \mathcal{L}_2 \cos \varphi-\mathcal{L}_1 \sin \varphi \ .
\end{equation}
Hence, the system is exactly solvable within an ansatz involving the $T$, $X$, $\xi$, $\phi$, $\varphi$, and $\chi$ coordinates. But the addition of dynamics in $\psi$ complicates matters significantly. 

We also need to assure that our ansatz is a consistent one: the first order equations must fit together and the second order geodesic equations must lead to valid trajectories. This step is computationally intensive, but at the end leads to two consistency conditions: one arising from $l_4^a$, and the other from $e^a$ 
\begin{equation}
	\label{eq:l4consistency} {\mathcal{L}_4}= -2 ({\mathcal{L}_1} \sin {\theta} \sin {\varphi}-{\mathcal{L}_2} \sin {\theta} \cos {\varphi}+{\mathcal{L}_3} \cos {\theta}) \ ,
\end{equation}
\begin{equation}
	\label{eq:energyconsistency} \mathcal{E}= \frac{2 ({\mathcal{L}_1} \sin {\theta} \sin {\varphi}-{\mathcal{L}_2} \sin {\theta} \cos {\varphi}+{\mathcal{L}_3} \cos {\theta})}{\xi^4} 
\end{equation}
where we used~(\ref{eq:chi})-(\ref{eq:norm}) in arriving at these expressions. If equation~(\ref{eq:l4consistency}) is to be satisfied, then~(\ref{eq:energyconsistency}) leads to a constant $\xi$ - unless both $\mathcal{L}_4$ and $\mathcal{E}$ are zero. Since a constant $\xi$ does not fit our needs for a bulk geodesic, we then need to require 
\begin{equation}
	\label{eq:lecondition} {\mathcal{L}_4}=0\ \ \ ,\ \ \ {\mathcal{E}}=0\ . 
\end{equation}
Otherwise, there are no spacelike geodesics for the given initial conditions. And thus this set of equations,~(\ref{eq:chi}) to~(\ref{eq:lecondition}), constitutes the system we need to explore to understand equal-time correlators in the PFT.

\vspace{0.25in} {\bf NOTES:} \vspace{0.125in}

While the constants $\mathcal{L}_i$, $\mathcal{E}$, and $\mathcal{P}$ appearing in the equations above are convenient to use in detailed computations, the physical quantities of interest correspond to attributes of a particle of mass $m$ whose field theory underlies the geodesic optical approximation. To recover the physical quantities, one needs to apply a rescaling of the constants of motion to insert back the factors of $m$, as well as to undo the coordinate rescalings performed earlier in~(\ref{eq:rescalings}) and the unit choice $R=1$. To clarify things, it helps to briefly summarize how to reestablish physical units from any expression: 
\begin{itemize}
	\item Note the length dimensions of the relevant variables: $\xi$, $T$, $X$, and $\tau$ have zero length dimension; the constants of motion written in italic script $\mathcal{L}_i$, $\mathcal{E}$, and $\mathcal{P}$ have length dimension one. 
	\item Insert in any given equation the appropriate powers of $R$ to make the relation consistent with the unit assignments described in the previous point. 
	\item Rescale the angular momenta as in $\mathcal{L}_i\rightarrow L_i/m$ where the non-calligraphic constants $L_i$ are physical angular momenta and are hence dimensionless. Apply also the rescaling $\mathcal{P}\rightarrow G^{1/3} \Delta\, P/m$ and $\mathcal{E}\rightarrow G^{1/3} \Delta\, E/m$ with $P$ and $E$ now having the proper length dimension of minus one. 
\end{itemize}
In the upcoming discussion, at the conclusion of any analysis we perform these reparameterizations to quote more transparent results.

Finally, let us note a simple yet subtle point. Spacelike geodesics do not carry physical momentum. The parameter $\mathcal{P}$ should be viewed as a parametrization of the extent of the geodesic in the $X$ direction at its endpoints. Hence, we will also replace $\mathcal{P}$ with $\Delta X$ at the end of every relevant computation.

\section{Non-locality through R-charge}\label{sec:nonlocality}

To read off the quantum numbers of an operator represented in a geodesic computation, we need to carefully consider the location of the holographic screen. The background spacetime carries angular momentum: a probe fixed at a point in the coordinate system of the metric~(\ref{eq:puffmetric}) is dragged by the spacetime and carries angular momentum. And our analysis of the convergence rate of null geodesic suggests that the holographic screen may be located at $\xi_H\neq 0$ when the endpoints are inserted at a value of $\psi\neq \pi/2$ (see equation~(\ref{eq:holographicscreen})). 

Consider a PFT operator insertion at some $\xi=\xi_c$ and arbitrary $\psi$. Denoting the four-velocity of such a stationary operator by $u^a$, we need $u_a u^a=-1$, $u^\xi=u^X=u^\psi=0$, and $u^T\neq 0$. The other angular components of $u^a$ must however be matched with those of the spacelike geodesic $v^a$ 
\begin{equation}
	\left.v^A\right|_{\xi_c}=\left.u^A\right|_{\xi_c}
\end{equation}
where $A$ is any of four angular directions $\phi$, $\varphi$, $\theta$, and $\chi$. This is so that the stationary operator insertion is accorded the same boundary conditions in the angles as the corresponding geodesic. We can then read off the R-charge by 
\begin{equation}\label{eq:qeq}
	\mathcal{Q}_i={l}_{i}^a u_a\ . 
\end{equation}
Hence, even when one has an insertion with no angular dynamics, we still would get $\mathcal{Q}_i\neq0$ because of the $g_{T\phi}$ term in the metric. This is the familiar frame dragging phenomenon that arises for example in Kerr black hole spacetimes when one asks about the angular momentum carried by a stationary observer. We will also see below that, for an operation insertion to correspond to a bulk geodesic computation satisfying the previous ansatz consistency conditions, equation~(\ref{eq:qeq}) will lead to an expression that cannot be made to vanish by fine tuning only the angular velocities of the insertion; instead, this is achieved if one also tunes the angular location.

From~(\ref{eq:volumeq4}), we expect that the R-charge that probes non-local effects is given by $\mathcal{Q}_4$. Applying the prescription just discussed to $\mathcal{Q}_4$, one gets 
\begin{eqnarray}
	&& \mathcal{Q}_4 = -\frac{\cos ^2\psi}{8 \xi ^5 \sqrt{\xi ^6+\cos ^2(\psi)}} \left[64 \mathcal{L}_0^2 \xi ^{10} \csc ^2\psi+256 \xi^{10} \sec ^2\psi \mathcal{L}^2 \right. \nonumber \\
	&& \left. + 64 \xi ^7 \sqrt{\xi ^6+\cos ^2\psi}+64 \xi ^4 \left(\xi ^6+\cos ^2\psi\right) \psi '(\tau )^2\right]^{1/2} 
\end{eqnarray}
where we define 
\begin{equation}
	\mathcal{L}^2\equiv{\mathcal{L}}^2_1+\mathcal{L}_2^2+\mathcal{L}_3^2 
\end{equation}
and we show the most general expression, including the effect of non-constant $\psi$. At $\xi=\xi_H$, this becomes 
\begin{equation}
	\label{eq:q4} \mathcal{Q}_4=-\sqrt{\frac{2}{3}} \sqrt{{\mathcal{L}_0}^2 \cot^2{\psi}+4 \mathcal{L}^2+\sqrt{3} \cos^2{\psi}} 
\end{equation}
where we now used $\psi'=0$. The only way to make this expression vanish is for $\mathcal{L}_1=\mathcal{L}_2=\mathcal{L}_3=0$ {\em and} $\psi\rightarrow \pi/2$ (even when ${\psi}'\neq 0$). Hence, if these conditions are satisfied, we should not expect non-local effects in correlation function computations. However, we also see that, when $\psi\rightarrow \pi/2$, $\mathcal{Q}_4$ can still be non-zero if $\mathcal{L}$ is non-zero. Looking more closely at the dynamics in the $\psi$ coordinates, we find that near $\psi\simeq 0$ 
\begin{equation}
	\psi''=\frac{\mathcal{L}_0^2\, \xi ^6 \cot \psi \csc ^2 \psi}{\xi^6+\cos ^2 \psi} 
\end{equation}
where we set $\psi'=0$; we see that $\psi=0=\mbox{constant}$ is a consistent ansatz if $\mathcal{L}_0=0$. Whereas near $\psi\rightarrow \pi/2$, we get 
\begin{equation}
	\psi''= -4 \sec ^3 \psi\, \mathcal{L}^2\ . 
\end{equation}
The $\psi\rightarrow \pi/2=\mbox{constant}$ is then a consistent ansatz only if $\mathcal{L}=0$. 

Hence, whenever $\mathcal{Q}_4\neq 0$ - and within the $\psi=\mbox{constant}$ ansatz - the consistent operator insertion is necessarily at $\psi=0\Rightarrow\xi_H=2^{-1/6}$, splitting the space in two. And $\mathcal{Q}_4\rightarrow 0$ as $\psi\rightarrow \pi/2$, moving the holographic screen $\xi_H\rightarrow 0$. Beside the $\psi=0$ and $\psi\rightarrow \pi/2$ cases, any other initial condition for $\psi$ leads to non-constant $\psi$ along the geodesic: a case that is significantly more complicated to analyze and that we explore numerically in Section~\ref{sec:psidynamics}. We then arrive at a remarkable conclusion: {\em whenever non-local effects are expected from PFT correlators that involve operators of the same charge, the associated operator insertions in the bulk spacetime necessarily involve a holographic screen splitting the space in two regions}; along with Bousso's criterion for holography, we are then to deal with a new realm of holography where two bulk spaces are projected onto one holographic screen.

\section{Local operators} \label{sub:local_operators}

In this section, we look at cases involving operators with zero or small $\mathcal{Q}_4$ R-charge, {\em i.e.} correlators which are not expected to exhibit non-local effects. The discussion in the previous section suggests that we need to insert the operators at $\psi$ near $\pi/2$. From our earlier analysis of the regime of validity, we also know that we can make $\mathcal{Q}_4$ parametrically small with larger values of the coupling $G$. Putting things together, we consider spacelike geodesics in a constant $\psi$ plane with $\psi$ arbitrarily close to $\pi/2$ with arbitrarily small $\mathcal{Q}_4$ R-charge; the endpoints of the geodesic are to be located at some small value $\xi=\xi_c$, corresponding to a large but finite UV cutoff in the dual PFT computation. To make the analysis more interesting, we will also turn on a non-zero value for $\mathcal{L}_0$. This still corresponds to vanishing $\mathcal{Q}_4$, but non-zero $\mathcal{Q}_0$. 

We start by looking at the equation of motion for $\psi$ (setting $\psi'=\psi''=0$) 
\begin{equation}
	\label{eq:psi1} {\sin \psi \cos \psi \left( \frac{2\,\xi^6 \mathcal{L}_0^2  \csc^4 \psi+2\,\xi^2 \mathcal{P}^2}{\xi ^6+\cos^2\psi}-\frac{\xi^3}{\left(\xi ^6+\cos ^2\psi\right)^{3/2}} \right)} = 0 
\end{equation}
confirming the $\psi\rightarrow \pi/2$ ansatz as a consistent one. We note that the initial condition in $\psi$ corresponds to fixing the R-charge of the boundary operator; a statement that is not subject to thermal or quantum fluctuations due to charge conservation. We also expect that this R-charge is most likely quantized in the full theory. In this scenario, the only non-zero or significant R-charge is 
\begin{equation}
	\mathcal{Q}_0=\mathcal{L}_0 \ .
\end{equation}

\begin{figure}
	\begin{center}
		\includegraphics[width=6in]{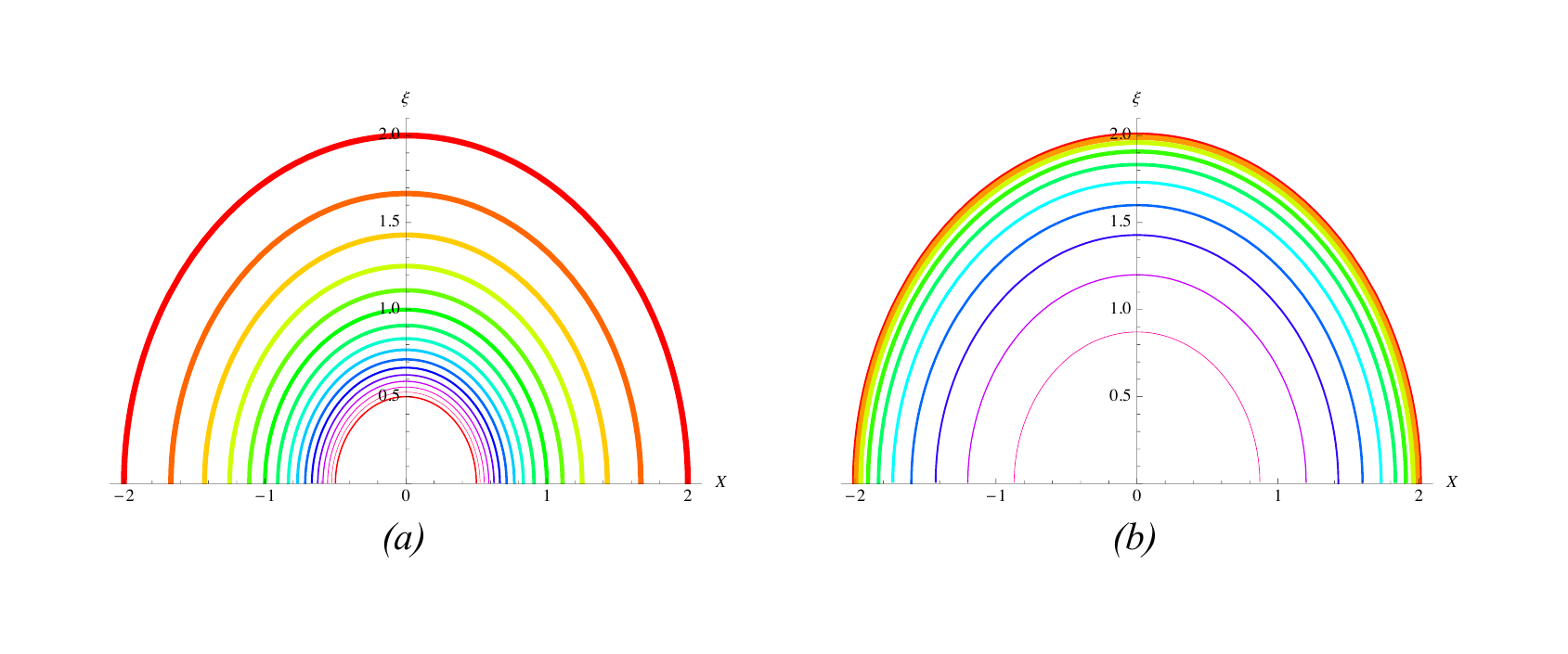} 
	\end{center}
	\caption{\em (a) Spacelike geodesics for varied values for $\mathcal{P}$ (and hence $\Delta X$). Thicker (redder) lines correspond to smaller values for $\mathcal{P}$. (b) Spacelike geodesics for fixed $\mathcal{P}$ and varied values for $\mathcal{L}_0$. Thicker (redder) lines correspond to smaller values for $\mathcal{L}_0$. At the maximum value of $\mathcal{L}_0=1$, the radius of the circle is zero.} \label{fig:geodesicforcase1} 
\end{figure}

The energy of the geodesic endpoints is given by 
\begin{equation}
	\mathcal{E}=\frac{\sqrt{1+{\mathcal{L}_0}^2}}{\xi_c}\Rightarrow E=\mu_c \sqrt{m^2 R^2+L_0^2} 
\end{equation}
where we have restored physical units as outlined at the end of Section~\ref{sec:geodesics} and used the UV-IR relation~(\ref{eq:uvir}). $E$ would presumably be related to the energy of the state associated with the given operator when applied to the vacuum - as measured in the PFT with respect to the original time coordinate $t$ . The conservation statements yield the first order equations 
\begin{equation}
	\chi'= {\mathcal{L}_0} \ \ \ , \ \ \ {X'}= \mathcal{P} \xi^2 \ \ \ , \ \ \ \xi'= \xi \sqrt{1-{\mathcal{L}_0}^2-\mathcal{P}^2 \xi^2} \ .
\end{equation}
The $\chi$ dynamics is trivial. The shape in the $\xi-X$ plane is slightly more interesting 
\begin{equation}
	\frac{d\xi}{dX}=\frac{\sqrt{1-{\mathcal{L}_0}^2-\mathcal{P}^2 \xi^2}}{\mathcal{P} \xi } \Rightarrow \xi_{cr}=\frac{\sqrt{1-\mathcal{L}_0^2}}{\mathcal{P}} 
\end{equation}
where $\xi_{cr}$ is the turning point of the geodesic - the maximum extent in the $\xi$ direction the geodesic explores the bulk. Integrating between $\xi_c\rightarrow 0$ and $\xi_{cr}$, we get 
\begin{equation}
	\Delta X = 2\, \frac{\sqrt{1-\mathcal{L}_0^2}}{\mathcal{P}} 
\end{equation}
where $\Delta X$ is the separation between the operator insertions in the PFT. The geodesics are simply circles 
\begin{equation}
	\frac{1-{\mathcal{L}_0}^2}{\mathcal{P}^2}=X^2+\xi^2 \ .
\end{equation}
And we note an upper bound on the angular momentum, and hence the R-charge 
\begin{equation}
	L_0\leq m R\ . 
\end{equation}
Otherwise, there are no spacelike geodesics connecting the operator insertions. Figure~\ref{fig:geodesicforcase1}(a) shows the shape of such geodesics as a function $\mathcal{P}$; these circles are common in AdS backgrounds. This is because our metric looks like AdS space as we approach the $\psi\rightarrow\pi/2$ plane. The effect of $\mathcal{L}_0$ is shown in Figure~\ref{fig:geodesicforcase1}(b). 

The length element of a geodesic is given by 
\begin{equation}
	dl=d\xi \sqrt{\frac{\mathcal{P}^2}{1-\mathcal{L}_0^2-\mathcal{P}^2 \xi ^2}+\frac{\mathcal{L}_0^2}{\xi^2 (1-\mathcal{L}_0^2-\mathcal{P}^2 \xi ^2)}+\frac{1}{\xi ^2}} 
\end{equation}
yielding a total length of
\begin{equation}
	l_{\xi_c} = -\left. {2 \ln \left(\frac{2 \left(\sqrt{{\mathcal{L}_0}^2+\mathcal{P}^2 \xi^2-1}-i \sqrt{1-{\mathcal{L}_0}^2}\right)}{\xi }\right)} \right|_{\xi_c}^{\frac{\sqrt{1-\mathcal{L}_0^2}}{\mathcal{P}}}\rightarrow  \frac{2}{\sqrt{1-\mathcal{L}_0^2}}\ln \frac{\Delta X}{\xi_c} 
\end{equation}
where we took the limit for the UV cutoff $\xi_c\rightarrow 0$, and we replaced $\mathcal{P}$ with $\Delta X$. This gives a correlation function of the form 
\begin{equation}
	e^{- m\, l_{\xi_c}} = \left( \frac{\xi_c}{\Delta X} \right)^{\frac{2\, m}{\sqrt{1-\mathcal{L}_0^2}}}\rightarrow \left( \frac{\mu_c^{-1}}{\Delta x} \right)^{\frac{2\, m\, R}{\sqrt{1-\mathcal{L}_0^2}}}
\end{equation}
after rescaling to physical variables. This is the 2-point correlator for a {\em scale invariant local} theory for an operator of dimension $2\, h_+ = m\, R/\sqrt{1-\mathcal{L}_0^2}$, as expected from~(\ref{eq:weight}). And all instances of $\mathcal{L}_0$ disappear. Note the limit $\mathcal{L}_0\rightarrow 1$ requires $m\rightarrow 0$: this case probably corresponds to a BPS configuration.

\section{Non-local operators} 

To explore non-local effects, our analysis of Section~\ref{sec:nonlocality} suggests that we need to insert the operators away from $\psi= \pi/2$. We consider spacelike geodesics in the $\psi=0$ plane, with endpoints at a tunable location $\xi=\xi_c$, corresponding to a UV cutoff in the dual PFT. In this case however, we entertain the possibility to move $\xi_c$ into the bulk away from $\xi_c\rightarrow 0$ limit since the holographic screen for $\psi=0$ is located at (see equation ~(\ref{eq:holographicscreen})) 
\begin{equation}
	\xi_H(\psi=0)=2^{-1/6}\equiv \xi_0 \ .
\end{equation}
We then have two qualitative scenarios: $\xi_c>\xi_0$ and $\xi_c<\xi_0$. In the former case, Bousso's criterion for holography identifies the region $\xi>\xi_c$ for holographic projection onto $\xi_c$; in the latter case, the region $\xi<\xi_c$ gets projected onto $\xi_c$. We will see below that the shape of spacelike geodesics lends itself to a natural interpretation of this novel holographic phenomenon.

Equation~(\ref{eq:q4}) suggests that we need to set $\mathcal{L}_0$ to zero when we take $\psi=0$. However, we can consider non-zero values for $\mathcal{L}_i$ with $i=1,2,3$. Note that $\mathcal{L}_4=0$ for a consistent ansatz as seen earlier from~(\ref{eq:l4consistency}). Looking at the second order geodesic equations, we find the following: 
\begin{itemize}
	\item $\mathcal{L}_{1,2}\neq 0$ activates the $\theta$ coordinate. 
	\item $\mathcal{L}_3\neq 0$ activates the $\phi$ and $\varphi$ coordinates. 
\end{itemize}
The $\psi$ equation of motion with $\psi'=\psi''=0$ leads to 
\begin{equation}
	\label{eq:psi2} 
	{\sin \psi \cos \psi \left( \frac{-8\, \xi ^6 \mathcal{L}_i^2\sec^4\psi+2\,\xi^2 \mathcal{P}^2}{\xi ^6+\cos^2\psi}-\frac{\xi^3}{\left(\xi ^6+\cos ^2\psi\right)^{3/2}} \right)} = 0 
\end{equation}
for any $i=1,2,3$. Hence, $\psi=0$ is a valid ansatz.

For simplicity, we consider all $\mathcal{L}_i$'s but $\mathcal{L}_3$ vanishing. The limit $\mathcal{L}_3\rightarrow 0$ was studied in detail in~\cite{Minton:2007fd}. For all values of $\mathcal{L}_3$, we have non-zero $Q_4$ R-charge as long as operators are inserted at the $\psi=0$ plane.

Hence, we consider spacelike geodesics with endpoints at $\xi=\xi_c>0$ and $\psi=0$; and $\mathcal{L}_3\neq 0$. The interesting R-charge becomes 
\begin{equation}
	\mathcal{Q}_4=-\frac{\sqrt{\xi_c}}{\sqrt[4]{1+\xi_c ^6}}\frac{1}{\xi_c^2} \sqrt{1+\frac{4 \mathcal{L}_3^2 \xi_c ^3}{\sqrt{1+\xi_c^6}}} \ .
\end{equation}
In this case, the limit $\xi_c\rightarrow 0$ is not useful: the holographic screen - where the UV completion of the theory is presumably located (in the given sector of R-charge) - is at $\xi_c=\xi_0=2^{-1/6}$. The covariant holographic criterion identifies two regions of the bulk space, on either sides of $\xi_0$; both projected onto the $\xi=\xi_0$ boundary. Evaluating $\mathcal{Q}_4$ at $\xi_c=\xi_0$, and after rescaling to physical units, the charge becomes 
\begin{equation}
	Q_4= -\frac{2\, m\, R}{\sqrt{3}} \sqrt{\frac{\sqrt{3}}{2}+\frac{2 L_3^2}{m^2 R^2}} \ ,
\end{equation}
yielding a volume for the corresponding excitation in the PFT from~(\ref{eq:volumeq4}) 
\begin{equation}
	\mbox{Vol}\sim m\, R\, \Delta^3\sqrt{\frac{\sqrt{3}}{2}+\frac{2 L_3^2}{m^2 R^2}} \ .
\end{equation}
We also get non-zero values for $\mathcal{Q}_1$, $\mathcal{Q}_2$, and $\mathcal{Q}_3$ which we will not need.

\begin{figure}
	\begin{center}
		\includegraphics[width=6in]{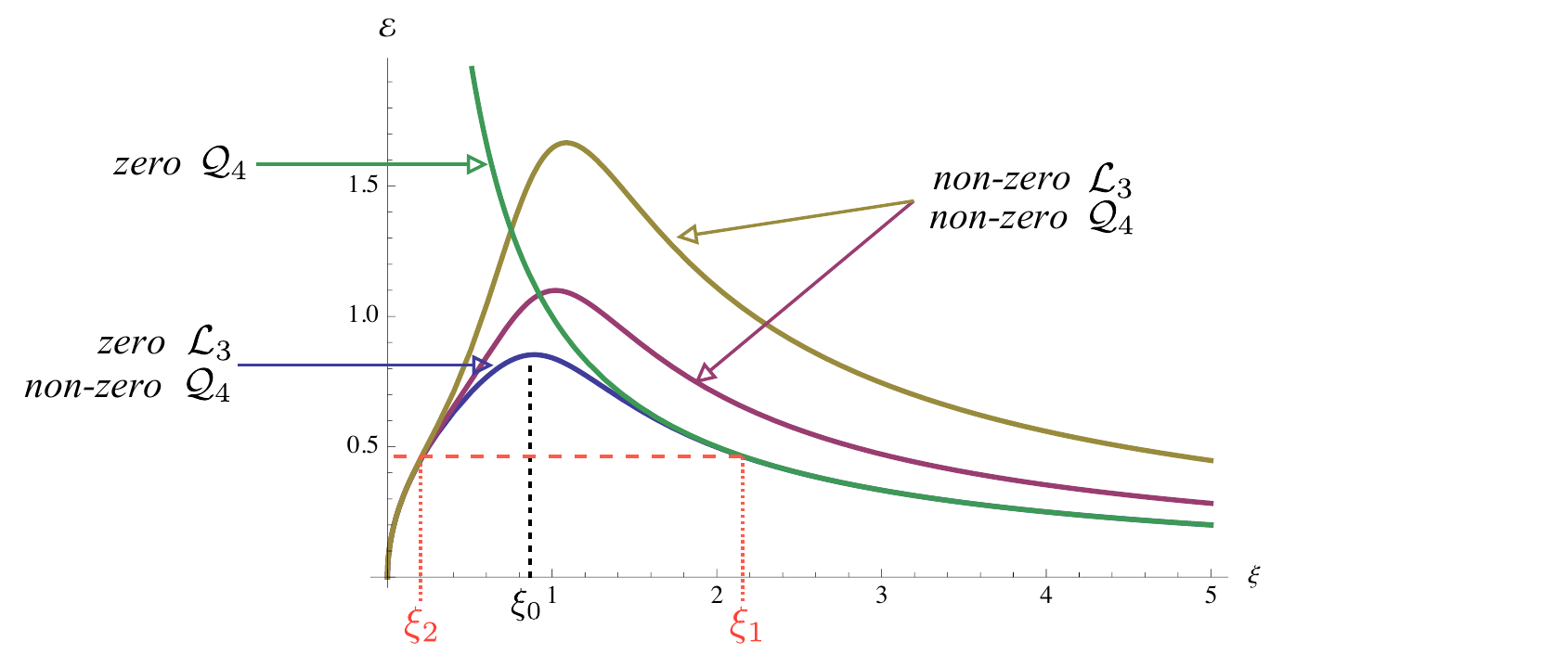} 
	\end{center}
	\caption{\em Plot of PFT energy for a state with non-zero $\mathcal{Q}_4$ charge contrasted with the equivalent expression for zero R-charge. The effect of increasing $\mathcal{L}_3$ (and hence $\mathcal{Q}_4$) is also shown.} \label{fig:energyplotforcase2} 
\end{figure}

The energy measured in the PFT for the given insertion is given by 
\begin{equation}\label{eq:uvir2}
	\mathcal{E}=\frac{\sqrt{\xi_c}}{\sqrt[4]{1+\xi^6_c}} \sqrt{1+\frac{{4 \mathcal{L}_i^2 \xi^3_c}}{\sqrt{1+\xi^6_c}}}\Rightarrow E=\mu_c \frac{m\, R}{\sqrt[4]{1+\frac{\mu^6}{\mu_{nl}^6}}} \sqrt{1+\frac{1}{\sqrt{1+\frac{\mu^6}{\mu_{nl}^6}}}\frac{4 L_3^2}{m^2 R^2}} \ .
\end{equation}
Figure~\ref{fig:energyplotforcase2} shows a plot of this expression. 
We note that, for $\mathcal{L}_3=0$, the maximum energy reading is attained when the cutoff is exactly at the holographic screen $\xi_c=\xi_0\Rightarrow \mu_c= 2^{1/6} \mu_{nl}$. As $\mathcal{L}_3$ is tuned up, the maximum shifts very slightly away from the holographic screen. An analytical treatment is not possible since it involves solving a polynomial of very high order. The folding shape of the energy profile will be important in helping us understand the holographic dictionary in the next section. 

\begin{figure}
	\begin{center}
		\includegraphics[width=6in]{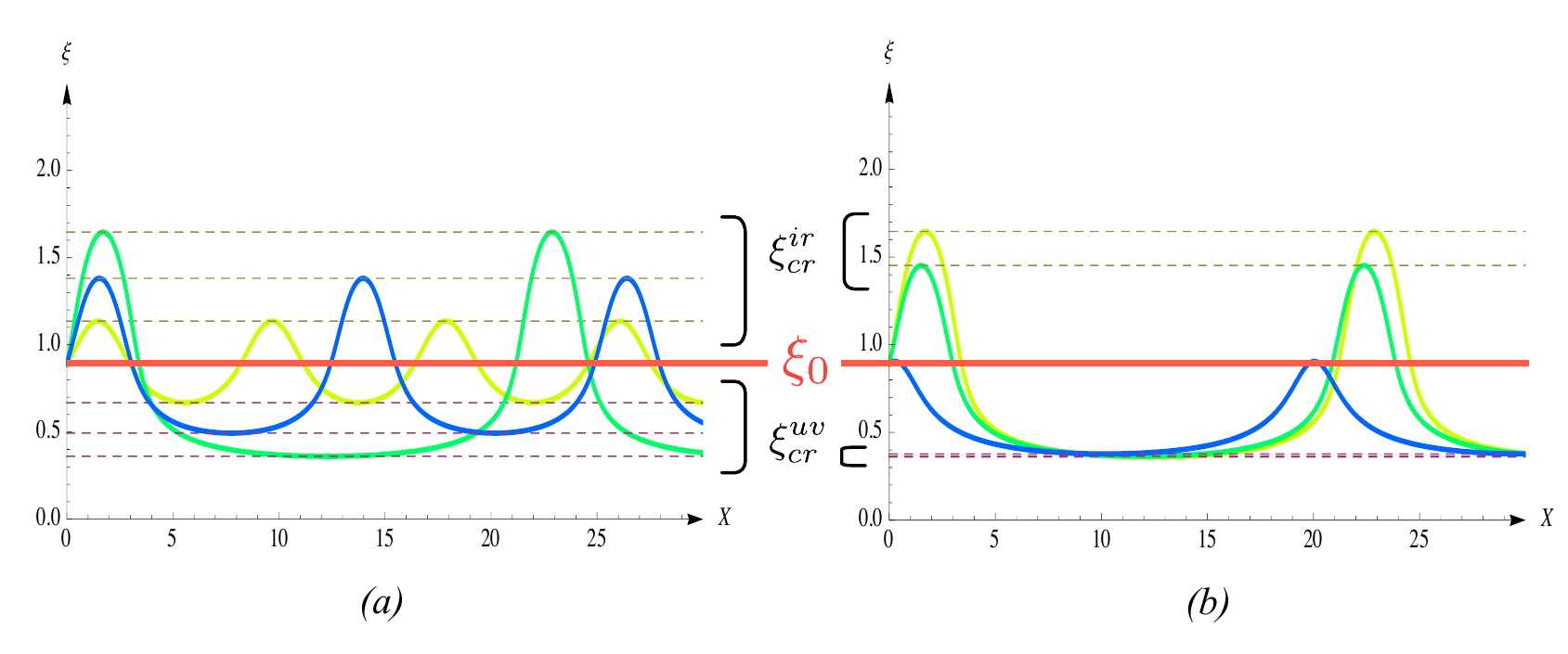} 
	\end{center}
	\caption{\em (a) Plots of geodesics with non-zero $\mathcal{Q}_4$ for $\mathcal{L}_3=0$ and varied values for $\mathcal{P}$. Curves with more oscillations correspond to larger values for $\mathcal{P}$; (b) Plots of geodesics with non-zero $\mathcal{Q}_4$ for increasing values of $\mathcal{L}_3$ at fixed $\mathcal{P}$. The curves shift towards the UV region as $\mathcal{L}_3$ is increased. At a critical value of $\mathcal{L}_3$ given by~(\ref{eq:bound}), the geodesic just grazes the holographic screen at $\xi=\xi_0$.} \label{fig:geodesicforcase2} 
\end{figure}

The conservation equations give us four first order differential equations 
\begin{equation}
	\phi'= -\frac{2 \, {\mathcal{L}_3} \xi^3}{\sqrt{1+\xi^6}}\ \ \ ,\ \ \ \varphi'= -\frac{4 \, {\mathcal{L}_3} \xi^3}{\sqrt{1+\xi^6}}\ , 
\end{equation}
\begin{equation}
	{X'}= \frac{\mathcal{P} \sqrt{1+\xi^6}}{\xi }\ \ \ ,\ \ \ \xi'= \frac{\xi^{2} \sqrt{{-4 {\mathcal{L}_3}^2 \xi^4-\mathcal{P}^2 (1+\xi^6)+\xi \sqrt{1+\xi^6} }}}{\left({1+\xi^6}\right)^{1/2}} \ .
\end{equation}
And hence we can write 
\begin{equation}
	\frac{d\xi}{dX}=\frac{\xi^{3} \sqrt{{-4 {\mathcal{L}_3}^2 \xi^4-\mathcal{P}^2 (1+\xi^6)+\xi \sqrt{1+\xi^6}\ . }}}{\mathcal{P}\left({1+\xi^6}\right)} 
\end{equation}
Writing $\xi_{cr}$ for the turning points of the geodesic, we find the condition 
\begin{equation}
	\label{eq:xicr} \frac{4 \mathcal{L}_3^2 \xi_{cr}^3}{\sqrt{\xi_{cr}^6+1}}+\frac{\mathcal{P}^2 \sqrt{\xi_{cr}^6+1}}{\xi_{cr}}=1 \ .
\end{equation}
Figure~\ref{fig:geodesicforcase2} shows a plot of the geodesics in the $\xi$-$T$ plane. 
We see that the geodesics oscillate about the holographic screen at $\xi=\xi_0$, between two critical points that we have labeled $\xi_{cr}^{ir}$ and $\xi_{cr}^{uv}$. These are the two real solutions of~(\ref{eq:xicr}). We call the region $\xi>\xi_0$ `the IR region'; and $\xi<\xi_0$ `the UV region'. As $\mathcal{L}_3$ is tuned up from zero, the geodesic's midpoint of oscillation shift in the $\xi$ direction from $\xi_0$ towards smaller values of $\xi$, towards the UV region. At a special point, the geodesic lies entirely in the region UV region. This occurs when the bound 
\begin{equation}
	\label{eq:bound} \frac{\mathcal{L}_3^2}{\mathcal{L}_m^2}+\frac{\mathcal{P}^2}{\mathcal{P}_m^2}\leq 1 
\end{equation}
is saturated, where we defined 
\begin{equation}
	\mathcal{L}_m\equiv\frac{\sqrt[4]{3}}{2}\ \ \ ,\ \ \ \mathcal{P}_m\equiv\frac{\sqrt[6]{2}}{\sqrt[4]{3}} \ .
\end{equation}
This can be determined from the condition $d\xi/d\tau=0$. If~(\ref{eq:bound}) is not satisfied, there are no spacelike geodesics with the given initial conditions connecting the operator insertions. For a given value of $\mathcal{L}_3$, (\ref{eq:bound}) translates to an upper bound on $\mathcal{P}$; for example, for $\mathcal{L}_3=0$, we get $\mathcal{P}<\mathcal{P}_m$. We will see later that this translates to a minimum bound on the separation between the operator insertions! 

\begin{figure}
	\begin{center}
		\includegraphics[width=6in]{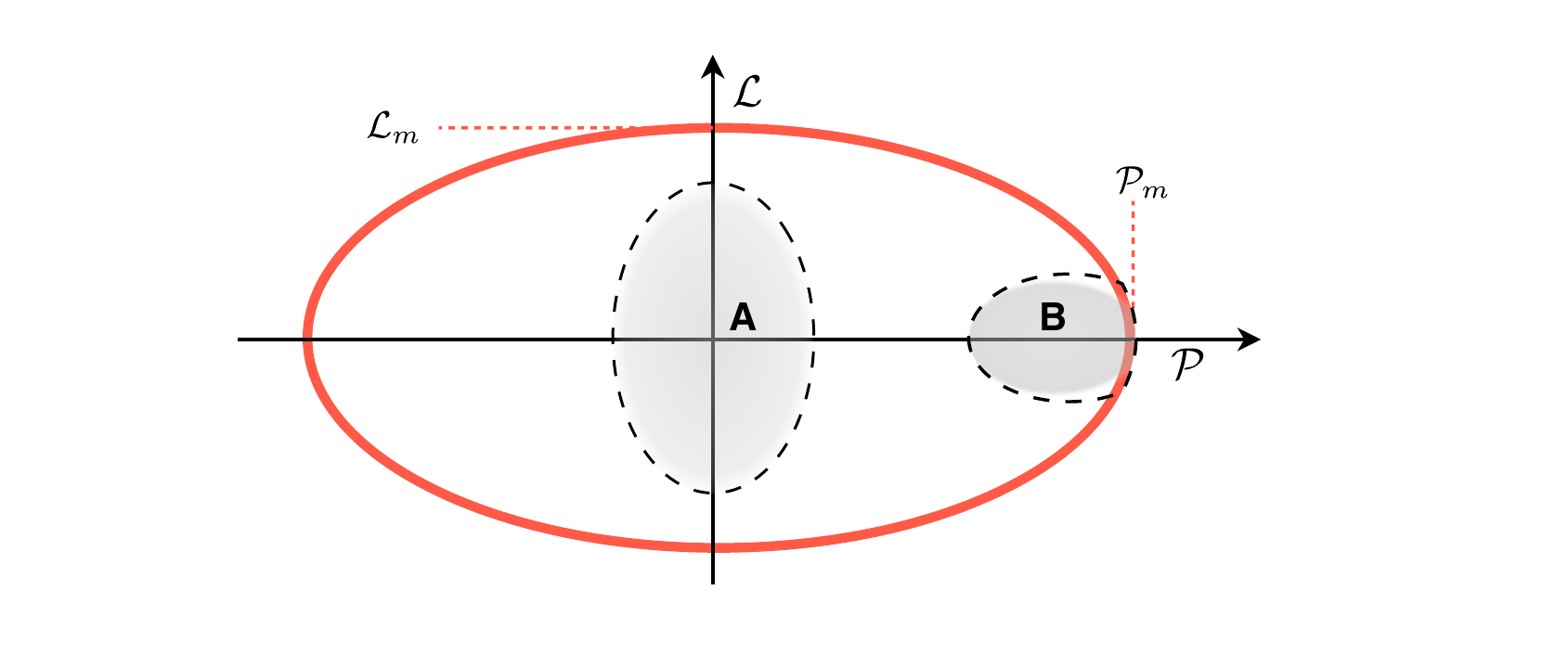} 
	\end{center}
	\caption{\em The parameter space of the geodesics used in the text. The region outside the ellipse is inaccessible with the chosen ansatz. Regions A and B correspond to regimes where asymptotic expansions of physical quantities are computed.} \label{fig:criticalellipse} 
\end{figure}

The geodesics will hence be parameterized by $\mathcal{P}$ and $\mathcal{L}_3$, and Figure~\ref{fig:criticalellipse} depicts this parameter space. 
Regions A and B denote regimes where we will be able to perform asymptotic expansions of various physical expressions. To see this, consider the coordinate change 
\begin{equation}
	\label{eq:trick} \lambda \equiv \frac{\xi }{\mathcal{P}_m^2 \sqrt{\xi ^6+1}} \ .
\end{equation}
We can then invert~(\ref{eq:trick}) in terms of $\lambda$ 
\begin{equation}
	\xi =\frac{\sqrt{\sqrt[3]{\sec (\Lambda )} \left(\cos \left(\frac{\Lambda }{3}\right)\mp\sqrt{3} \sin \left(\frac{\Lambda }{3}\right)\right)}}{\sqrt[6]{2}} 
\end{equation}
where 
\begin{equation}
	\lambda = \frac{1}{\sqrt[6]{\tan ^2(\Lambda )+1}}\ . 
\end{equation}
This allows us to consider the following two regimes: small $\lambda$ and $\lambda$ near unity. For small $\lambda$, equation~(\ref{eq:xicr}) for the turning points implies that we must have $\mathcal{P}\ll \mathcal{P}_m$. To further simplify otherwise very cumbersome analytical expressions, we will also consider $\mathcal{L}_3\ll \mathcal{L}_m$. These statements lead to region A in Figure~\ref{fig:criticalellipse}. For $\lambda$ near unity, (\ref{eq:xicr}) asymptotes to~(\ref{eq:bound}), or equivalently $\mathcal{P}$ and $\mathcal{L}_3$ are such that we are saturating~(\ref{eq:bound}) {\em and} $\mathcal{P}\sim \mathcal{P}_m$ since both turning points are also approaching $\xi_0$: that is, we explore region B in Figure~\ref{fig:criticalellipse}. 

We now can write asymptotic forms for the turning points of the geodesics:
\begin{equation}
	\xi^{ir}_{cr} = \left\{ 
	\begin{array}{ll}
		\frac{\sqrt{1-4 \mathcal{L}_3^2}}{\mathcal{P}} & \mathcal{P}\ll \mathcal{P}_m\ ,\ \mathcal{L}_3\ll \mathcal{L}_m \\
		{\xi_0 \left ( 1 + \frac{1}{\sqrt{2}}\sqrt{1 - \frac{\mathcal{P}^2}{\mathcal{P}_{m}^2}-\frac{\mathcal{L}_3^2}{\mathcal{L}_m^2}} -\frac{1}{2} \frac{\mathcal{L}_3^2}{\mathcal{L}_m^2}\right )} & \mathcal{P}\,\widetilde{<}\, \mathcal{P}_m\ ,\ \mathcal{L}_3\ll \mathcal{L}_m 
	\end{array}
	\right. 
\end{equation}
\begin{equation}
	\xi^{uv}_{cr} = \left\{ 
	\begin{array}{ll}
		\mathcal{P}^2 \sqrt[3]{1+12 \mathcal{L}_3^2\, \mathcal{P}^6} & \mathcal{P}\ll \mathcal{P}_m\ ,\ \mathcal{L}_3\ll \mathcal{L}_m \\
			{\xi_0 \left ( 1 - \frac{1}{\sqrt{2}}\sqrt{1 - \frac{\mathcal{P}^2}{\mathcal{P}_{m}^2}-\frac{\mathcal{L}_3^2}{\mathcal{L}_m^2}} -\frac{1}{2} \frac{\mathcal{L}_3^2}{\mathcal{L}_m^2}\right )} & \mathcal{P}\,\widetilde{<}\, \mathcal{P}_m\ ,\ \mathcal{L}_3\ll \mathcal{L}_m 
	\end{array}
	\right. 
\end{equation}

\begin{figure}
	\begin{center}
		\includegraphics[width=6in]{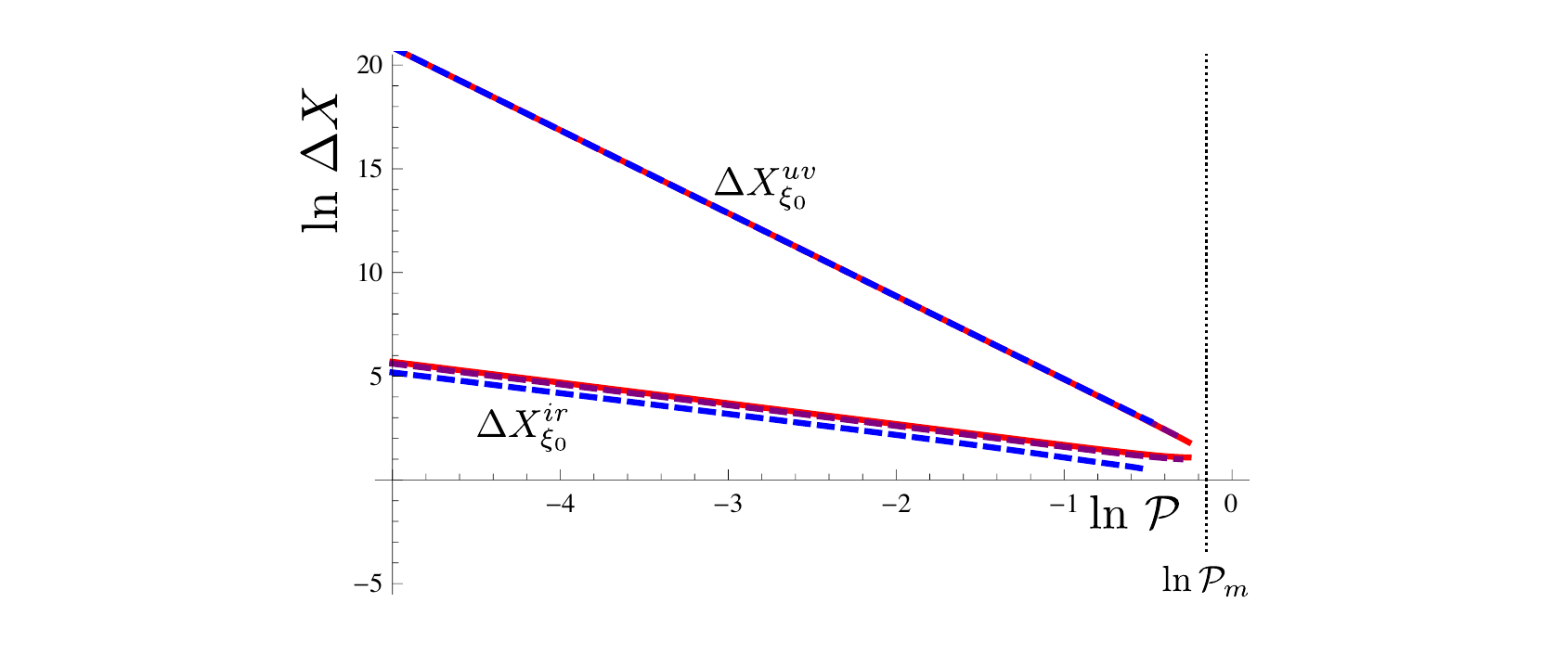} 
	\end{center}
	\caption{\em A log plot of $\Delta X$ versus $\mathcal{P}$. The solid lines correspond to the $\mathcal{L}_3=0$ case; the dashed lines demonstrate the effect of increasing $\mathcal{L}_3$ to $0.2$ and $0.4$ (where $\mathcal{L}_m$ is about $0.65$). } \label{fig:dispersionplotcase2} 
\end{figure}

To compute the correlator, we would need the length of the geodesic 
\begin{equation}
	l_{\xi_c}^{ir,uv} = 2\, R \int_{\xi_c}^{\xi^{ir,uv}_{cr}} \frac{d\xi}{\xi^2} \sqrt{\frac{\left(1+\xi ^6\right)}{\left(-4 \mathcal{L}_3^2\, \xi ^4-\mathcal{P}^2(1 + \xi ^6)+\xi\, \sqrt{1+\xi ^6} \right)}} \ .
\end{equation}
To simplify the discussion, we henceforth move the cutoff plane $\xi_c$ to $\xi_0$. We then get the asymptotic expressions 
\begin{equation}
	l^{ir}_{\xi_0} = R \left\{ 
	\begin{array}{ll}
		\frac{2}{{1-2 \mathcal{L}_3^2}} \ln \left(\frac{1-2 \mathcal{L}_3^2}{\mathcal{P}}\right) & \mathcal{P}\ll \mathcal{P}_m\ ,\ \mathcal{L}_3\ll \mathcal{L}_m \\
		\frac{2^{2/3}}{\mathcal{P}_m} \left( 1-\frac{1}{\sqrt{2}}\sqrt{1-\frac{\mathcal{P}^2}{\mathcal{P}_m^2}-\frac{\mathcal{L}_3^2}{\mathcal{L}^2_m}}  - \frac{1}{2\sqrt{2}} \frac{\mathcal{L}_3^2}{\mathcal{L}^2_m} \left( 1-\frac{\mathcal{P}^2}{\mathcal{P}_m^2}-\frac{\mathcal{L}_3^2}{\mathcal{L}^2_m} \right)^{-1/2}\right) & \mathcal{P}\,\widetilde{<}\, \mathcal{P}_m\ ,\ \mathcal{L}_3\ll \mathcal{L}_m 
	\end{array}
	\right. 
\end{equation}
\begin{equation}
	l^{uv}_{\xi_0} = R \left\{ 
	\begin{array}{ll}
		\frac{26}{15 \mathcal{P}^3} +\frac{2^{3/4}}{\sqrt{3}} \frac{\mathcal{L}_3^2}{\mathcal{L}_m^2} & \mathcal{P}\ll \mathcal{P}_m\ ,\ \mathcal{L}_3\ll \mathcal{L}_m \\
		\frac{2^{2/3}}{\mathcal{P}_m} \left( 1+\frac{1}{\sqrt{2}}\sqrt{1-\frac{\mathcal{P}^2}{\mathcal{P}_m^2}-\frac{\mathcal{L}_3^2}{\mathcal{L}^2_m}}  + \frac{1}{2\sqrt{2}} \frac{\mathcal{L}_3^2}{\mathcal{L}^2_m} \left( 1-\frac{\mathcal{P}^2}{\mathcal{P}_m^2}-\frac{\mathcal{L}_3^2}{\mathcal{L}^2_m} \right)^{-1/2}\right) & \mathcal{P}\,\widetilde{<}\, \mathcal{P}_m\ ,\ \mathcal{L}_3\ll \mathcal{L}_m 
	\end{array}
	\right. 
\end{equation}
where $l^{ir}_{\xi_0}$ is the length of the geodesic folding into the $\xi>\xi_0$ region, and $l^{uv}_{\xi_0}$ is the length of the geodesic folding into the $\xi<\xi_0$ region.
To make sense of these expressions, we need to replace $\mathcal{P}$ with the separation $\Delta X$ between the endpoints of the geodesic. The relation between $\Delta X$ and $\mathcal{P}$ is given by
\begin{equation}\label{eq:deltax}
	\Delta X^{ir,uv}_{\xi_0} = 2\,\left| \int_{\xi_0}^{\xi_{cr}^{ir,uv}} d\xi\frac{\mathcal{P} (1+\xi^6)^{3/4}}{\xi^3 (\xi-\mathcal{P}^2 \sqrt{1+\xi^6}-4\,\mathcal{L}_3^2 \frac{\xi^4}{\sqrt{1+\xi^6}})^{1/2}} \right| \ ,
\end{equation}
Figure~\ref{fig:dispersionplotcase2} shows a plot of this relation, 
We note in particular a complication arising from the fact that this relation is not single-valued: for a given $\mathcal{P}$, we have two possible values for $\Delta X$ - one for the geodesic in the IR region, and another for the one in the UV region. $\Delta X^{ir}_{\xi_0}$ is a standard dispersion relation in a local theory, except that we have an upper bound on $\mathcal{P}$. And we see that changing $\mathcal{L}_3$ does not change these profiles much. Indeed, the small shifts we notice are within numerical errors. We also note that higher values of $\mathcal{L}_3$ bring down the maximum $\mathcal{P}$ bound, as expected from~(\ref{eq:bound}). The asymptotic forms of interest are 
\begin{equation}\label{eq:deltaxir}
	\Delta X^{ir}_{\xi_0}\simeq\left\{ 
	\begin{array}{ll}
		2 \frac{\sqrt{1-2\,\mathcal{L}_3^2}}{\mathcal{P}} & \mathcal{P}\ll \mathcal{P}_{m}\ ,\ \mathcal{L}_3\ll \mathcal{L}_m \\
		\frac{l^{ir}_{\xi_0}}{\mathcal{P}_m} & \mathcal{P}\,\widetilde{<}\, \mathcal{P}_{m}\ ,\ \mathcal{L}_3\ll \mathcal{L}_m 
	\end{array}
	\right. 
\end{equation}
and 
\begin{equation}\label{eq:deltaxuv}
	\Delta X^{uv}_{\xi_0}\simeq\left\{ 
	\begin{array}{ll}
		\frac{3\pi}{4} \mathcal{P}^{-4} + C_1 \mathcal{L}_3^2\mathcal{P} & \mathcal{P}\ll \mathcal{P}_{m}\ ,\ \mathcal{L}_3\ll \mathcal{L}_m \\
		\frac{l^{uv}_{\xi_0}}{\mathcal{P}_m}  & \mathcal{P}\,\widetilde{<}\, \mathcal{P}_{m}\ ,\ \mathcal{L}_3\ll \mathcal{L}_m 
	\end{array}
	\right. , 
\end{equation}
where $C_1\equiv {3\,2^{-1/12}\, {}_2\text{F}_1\left(\frac{1}{12},\frac{3}{4},\frac{13}{12},-\frac{1}{2}\right)}+{3^{1/4}}{2^{-1/3}}\simeq 4$. We note that we see $\Delta X\sim l$ in the $\mathcal{L}\ll\mathcal{L}_m$, $\mathcal{P}\,\widetilde{<}\,\mathcal{P}_m$ regime since the geodesics are asymptotically approaching the holographic screen. We have checked our asymptotic expansions against the numerical results: in the case of the $\mathcal{L}\ll\mathcal{L}_m$, $\mathcal{P}\ll\mathcal{P}_m$ regime, we find excellent agreement; for the $\mathcal{L}\ll\mathcal{L}_m$, $\mathcal{P}\,\widetilde{<}\,\mathcal{P}_m$ regime, the comparisons are complicated by numerical artifacts and convergence issues of integrals, yet the asymptotic expansions seem to match adequately with the numerical results.

We can then identify a minimum separation between the endpoints of the geodesics
\begin{equation}
	\Delta X_0\equiv \frac{2^{2/3}}{\mathcal{P}_m^2}\Rightarrow \Delta x_0= 2^{1/3} 3^{1/2} \mu_{nl}^{-1}\sim \mu_{nl}^{-1}\ ,
\end{equation}
which can be obtained from~(\ref{eq:deltaxir}) or~(\ref{eq:deltaxuv}) evaluated at $\mathcal{P}=\mathcal{P}_m$ and $\mathcal{L}_3=0$.

\begin{figure}
	\begin{center}
		\includegraphics[width=5.5in]{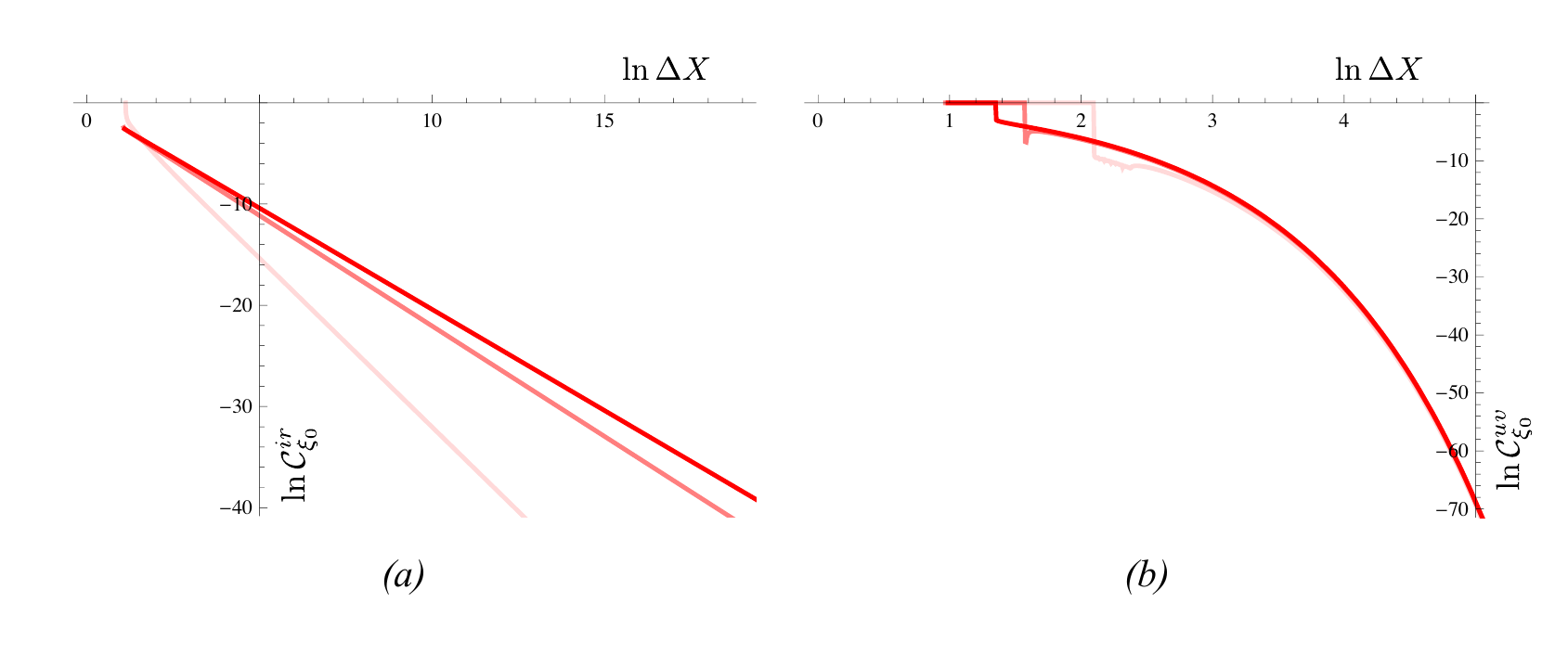} 
	\end{center}
	\caption{\em (a) Log plot of the correlator factor $\mathcal{C}_{\xi_0}^{ir}$ versus $\Delta X$ with varied values for $\mathcal{L}_3=0,0.2,0.4$; fainter lines have larger values of $\mathcal{L}_3$; (b) Same plot for $\mathcal{C}_{\xi_0}^{uv}$. Numerical artifacts arise at the minimum separation $\Delta X_m$.} \label{fig:correlatorplotcase2} 
\end{figure}

We finally put all these results together into building blocks for correlators: 
\begin{equation}
	\mathcal{C}_{\xi_0}^{ir}\equiv e^{-m\, l^{ir}_{\xi_0}}\simeq \left\{ 
	\begin{array}{ll}
		\Delta x^{-\frac{2\, m\, R}{{1-2 \mathcal{L}_3^2}}} \left( 1+2\, \frac{L_3^2}{m\, R} \right) & \Delta x \gg \Delta x_m \\
		e^{-m\, R\, \left( \frac{\Delta x}{\Delta x_m} \right)} & \Delta x \sim \Delta x_m 
	\end{array}
	\right. \ , 
\end{equation}
where $L_m\equiv 3^{1/8}\, 2^{-1/4}\,m\, R$; and 
\begin{equation}
	\mathcal{C}_{\xi_0}^{uv}\equiv e^{-m\, l^{uv}_{\xi_0}}\simeq \left\{ 
	\begin{array}{ll}
		e^{-m\, R\, \left( C_2 \frac{\Delta x^{3/4}}{\Delta x_m^{3/4}}+C_3 \frac{L_3^2}{L_m^2}\right)} & \Delta x \gg \Delta x_m \\
		e^{-m\, R\, \left( \frac{\Delta x}{\Delta x_m}\right)} & \Delta x \sim \Delta x_m 
	\end{array}
	\right. \ , 
\end{equation}
where $C_2 \equiv 52\, 2^{-3/8} 3^{-25/16} 5^{-1} \pi^{-3/4}\simeq 1$ and $C_3 \equiv {2^{3/4}}{3}^{-1/2}\simeq 1$; and we have defined
\begin{equation}
	\Delta X_m\equiv \mathcal{P}_m^{-1}\Rightarrow \Delta x_m= 3^{1/4} 2^{-1/6}\mu_{nl}^{-1}\sim \mu_{nl}^{-1}\ . 
\end{equation}
Once again, the characteristic scale of non-locality in the theory is then set by $\mu_{nl}^{-1}$.
Figure~\ref{fig:correlatorplotcase2} shows a plot of $\mathcal{C}_{\xi_0}^{ir}$ and $\mathcal{C}_{\xi_0}^{uv}$. 
We note in particular the sensitivity of the power law on $\mathcal{L}_3$ for $\mathcal{C}_{\xi_0}^{ir}$; whereas $\mathcal{C}_{\xi_0}^{uv}$ is mostly unaffected except for the shift in the value for the minimum operator separation. Identifying a weight for the operator in the $\Delta x\gg \Delta x_m$ regime, we write
\begin{equation}
	h_+\simeq\frac{m\, R}{2} \left( 1+\frac{2\, L_3^2}{m^2 R^2} \right) \ .
\end{equation}
We also note that the bound on $L_3$ translates to a bound on $Q_4$: as $\mathcal{L}_3$ varies from $0$ to $\mathcal{L}_m$, $Q_4$ varies from $-\sqrt{2}\,m\,R/\sqrt[4]{3}$ to $-\sqrt{2}\sqrt[4]{3}\,m\,R$, increasing by a factor of $\sqrt{3}$.

\section{Dynamics of the $\psi$ coordinate} \label{sec:psidynamics}

In this section, we briefly consider initial conditions for $\psi=\psi_0\neq 0$ or $\pi/2$, leading to changing $\psi$ along the geodesics. These trajectories necessarily correspond to R-charged non-local operator insertions. The equation of motion for $\psi$ is given by
\begin{eqnarray}
	&\psi''-\frac{3\, \psi'\,\cos ^2\psi}{\left(\xi ^6+\cos ^2\psi\right)^{3/4}}\sqrt{\xi ^3-\mathcal{P}^2 \xi ^2 \sqrt{\xi ^6+\cos ^2\psi}-{\psi'}^2 \sqrt{\xi ^6+\cos ^2\psi}} \nonumber \\
	&+\frac{1}{2} \sin \psi \cos \psi \left(\frac{\xi ^3}{\left(\xi ^6+\cos ^2\psi\right)^{3/2}}-\frac{2 \left(\mathcal{P}^2 \xi ^2+\psi'^2\right)}{\xi ^6+\cos ^2\psi}\right)=0
	\end{eqnarray}
For simplicity, we are considering spacelike geodesics in the $\xi$-$\psi$-$X$ plane, fixing all other angular directions to constants. This system cannot be solved in closed form due to lack of sufficient symmetry. Hence, we will contend with a brief numerical analysis. 

\begin{figure}
	\begin{center}
		\includegraphics[width=6in]{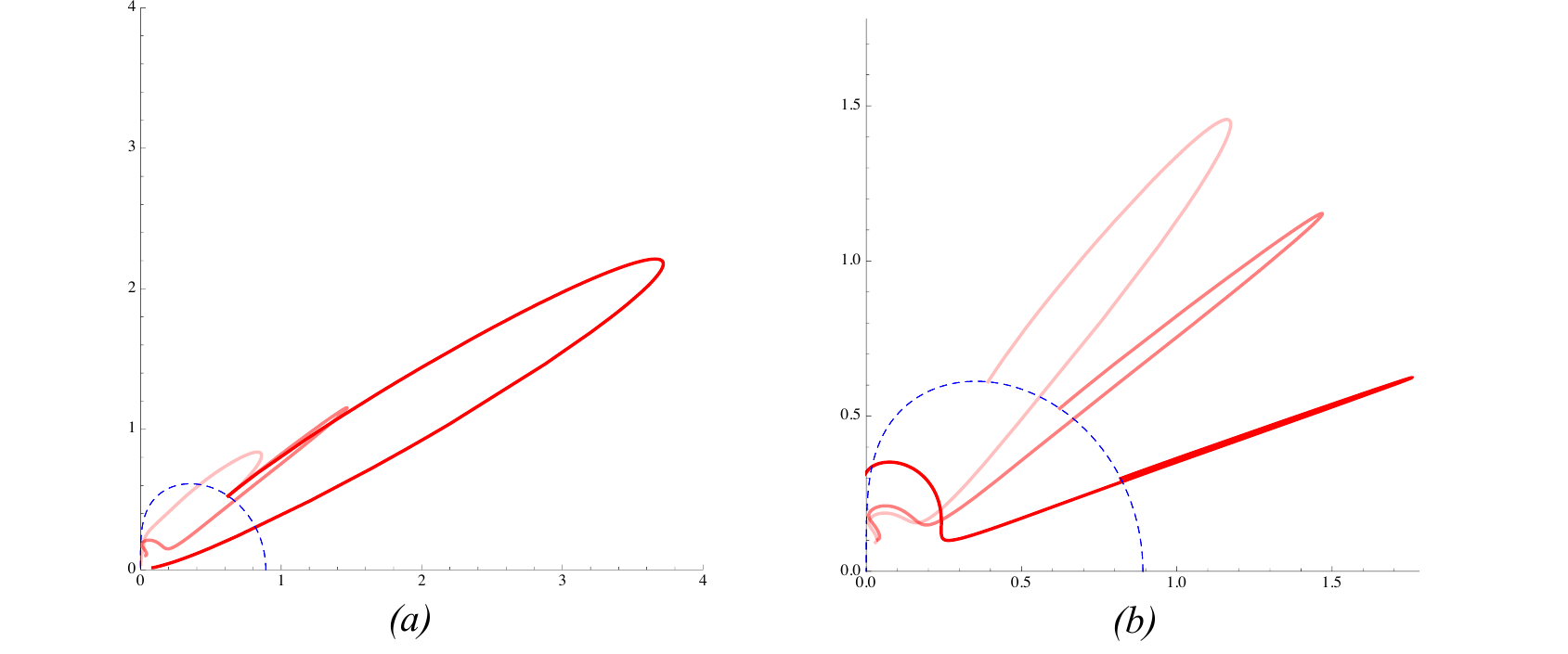} 
	\end{center}
	\caption{\em Shape of spacelike geodesics in the $\xi$-$\psi$ plane: $\xi$ is the radial direction, and $\psi$ is the polar angle. The dotted curve denotes the location of the holographic screen $\xi_H$. The red solid lines with increasingly fainter tones corresponds to: (a) varied $\mathcal{P}=1.3, 3, 4.5$ for fixed $\psi_0=0.7$; (b) varied $\psi_0=0.35, 0.7, 1.0$ for fixed $\mathcal{P}=3$.} \label{fig:psiplot} 
\end{figure}

Figure~\ref{fig:psiplot} shows plots of the shape of the geodesics. We first notice that the curve generically crosses the holographic screen as in the case when $\psi_0=0$. In \ref{fig:psiplot}(a), we see the effect of varying $\mathcal{P}$: we note a transition from attraction towards the $\psi=0$ to the $\psi=\pi/2$ at a critical $\mathcal{P}$, or a critical separation between the operator insertions locations. This implies that for certain separations between the insertions, there may be a non-zero correlation between non-local and local operators. In \ref{fig:psiplot}(b), we explore the effect of varying $\psi_0$: we note no particular interesting behavior in this regard. 
Given the complexity of the problem at hand, we will delay any further exploration and leave these numerical observations as they are.

\section{UV-IR mixing} \label{sec:mechanism_of_non_locality}

In this section, we try to make sense of the unusual holographic setup we have on our hands. Let us start by summarizing the main points that will drive us towards a new extended dictionary for holography. The holographic dual spacetime has a product  $\mathcal{X}\times \mathcal{M}$ structure, where $\mathcal{X}$ asymptotes to $AdS_5$ in the IR (large $\xi$); and $\mathcal{M}$ is a five dimensional compact space.

\begin{itemize}
	\item We showed, using Bousso's covariant criterion for holography, that the holographic screen gets shifted from the boundary  ({\em i.e. away from $\xi=0$}) to inside the bulk - as a function of an angle coordinate $\psi$ of $\mathcal{M}$: as $\psi$ is tuned from $\pi/2$ to zero, the holographic screen moves from $\xi=0$ at the boundary of the bulk to $\xi=\xi_0=2^{-1/6}$. Operators inserted at this new holographic screen sit at the boundary of two regions of the bulk - both holographically encoded onto their common boundary at $\xi=\xi_0$.
	\item From general and independent considerations, we expect non-local behavior from states that carry  particular R-charges. We showed conclusively that such states are realized in the holographic bulk {\em only if} the operators are inserted at $\psi\neq \pi/2$ - due to the angular momentum carried by the bulk spacetime. This matches with the condition that shifts the holographic screen into the bulk: $\psi$ tunes the value of the interesting R-charge {\em and} the location of the holographic screen.
	\item We showed that the new holographic screen corresponds to a minimal volume surface (in $x_1$-$x_2$-$x_3$) - the surface where the rate of convergence of a congruence of null geodesics vanishes. We know from previous work~\cite{Ryu:2006bv}-\cite{Nishioka:2009un} that such surfaces, when arranged to split a bulk spacetime by hand, generate quantum entanglement of states in the dual theory: states dual to one side of the surface are entangled with states on the other side.
	\item The thermodynamics of the PFT determines a UV-IR relation between bulk extent and energy scale in the PFT that seems insensitive to non-local effects. In fact, it is identical to the one arising in $\mathcal{N}=4$ SYM theory: small $\xi$ is in the UV, large $\xi$ in the IR.
	\item For states which are not expected to show non-local effects, we demonstrated that this was indeed the case and the scenario was reminiscent to the usual $\mathcal{N}=4$ SYM setup with the holographic screen at the boundary $\xi=0$.
\end{itemize}

\begin{figure}
	\begin{center}
		\includegraphics[width=6in]{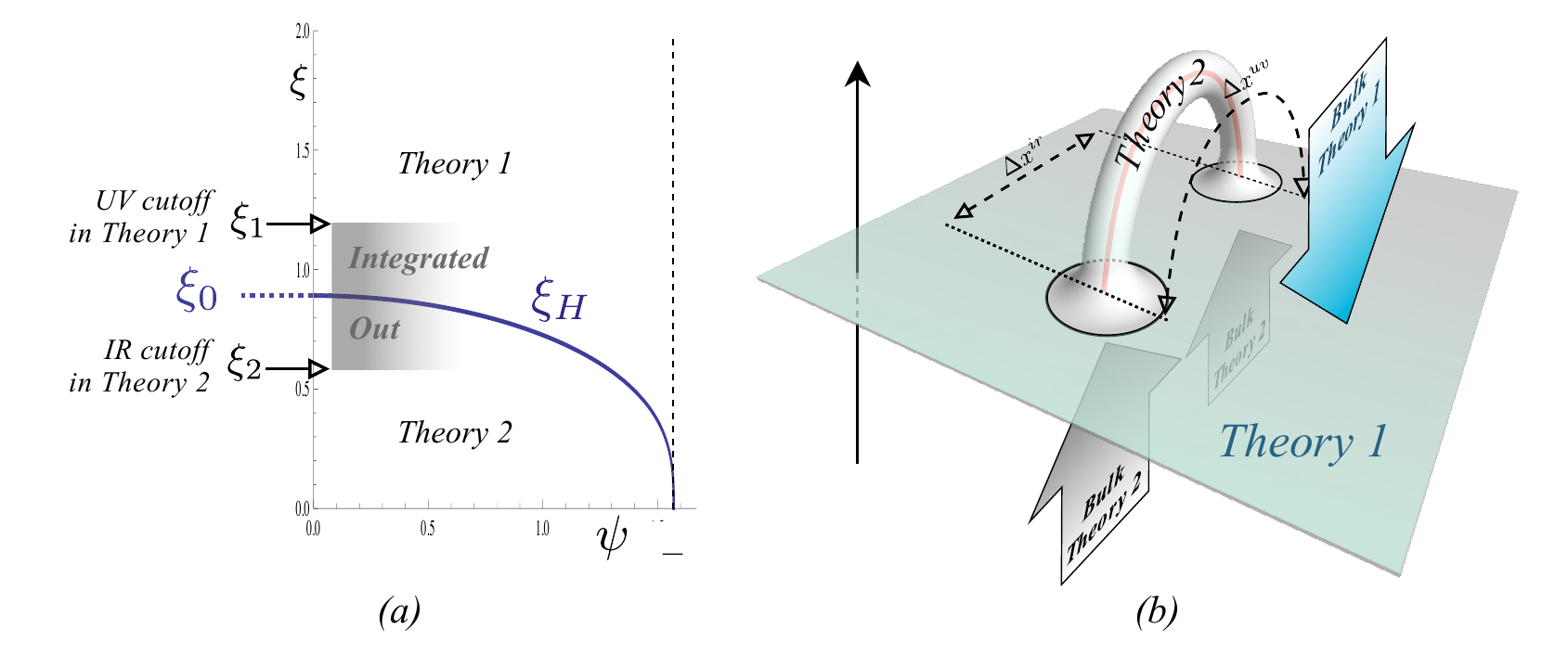} 
	\end{center}
	\caption{\em A new holographic dictionary deciphered: (a) A plot of the $\xi$-$\psi$ cross section; non-local operators are inserted at $\psi=0$ and $\xi=\xi_0$; (b) The two non-local operator insertions in the PFT amount to a D3 brane tentacle extending between them. Theory 1 computes the correlator for a separation of $\Delta x^{ir}$, with the insertions corresponding to deleted 3-spheres of volume $\sim\Delta x_m^3$. Theory 2 compute the correlator for separation $\Delta x^{uv}$ in the theory of the tentacle D3 brane.} \label{fig:entanglement} 
\end{figure}

Figure~\ref{fig:entanglement}(a) shows the $\xi$-$\psi$ cross section of the bulk. Non-local operators are inserted at $\psi=0$ and $\xi=\xi_H=\xi_0$; local operators are inserted near $\psi=\pi/2$ and $\xi=\xi_H=0$. Let's focus on the interesting $\psi=0$ case from hereon. If we were to employ a UV cutoff $\mu_1$, we would cut the bulk space at $\xi=\xi_1=\mu_{nl}/\mu_1$: the $\xi>\xi_1$ region is expected to be projectable onto the $\xi=\xi_1$ boundary by the covariant holography criterion. 
We would compute the corresponding PFT 2-point correlator through the length of a geodesic folding into the $\xi>\xi_1$ region. We can then push the UV cutoff to the maximum value $\mu_1\rightarrow\mu_{0}\equiv \mu_{nl}/\xi_0$. We previously labeled this geodesic length as $l_{\xi_0}^{ir}$. We then have
\begin{equation}\label{eq:pftcorr1}
	\langle \mathcal{O}(x_1) \mathcal{O}(x_2) \rangle_{\mu_{0}} \simeq \left\{ 
	\begin{array}{ll}
		\left( \Delta x \right)^{-\frac{2\, m\, R}{{1-2 \mathcal{L}_3^2}}}  & \Delta x \gg \Delta x_m \\
		e^{-m\, R\, \left( \frac{\Delta x}{\Delta x_m}\right)} & \Delta x \sim \Delta x_m 
	\end{array}
	\right. \ 
\end{equation} 
This result, which is the leading tree level expression for the correlator, interpolates between a scale invariant behavior and one with characteristic non-local effects. The scale of non-locality is set by $\Delta x_m\sim \Delta x_0\sim \mu_{nl}^{-1}$ - the minimum distance between the two operator insertions. If the operators are viewed as puffed up D3 brane bubbles, the volume of these spherical states is roughly $\Delta x_m^3\sim G\,\Delta^3$.

The holographic screen at $\xi=\xi_0$ is also a surface of minimum volume, a neck in the global geometry of the bulk. From independent considerations~\cite{Ryu:2006bv}-\cite{Nishioka:2009un}, we then expect that~(\ref{eq:pftcorr1}) involves entanglement effects from integrating out states behind the holographic screen, the region $\xi<\xi_0$. Such an entanglement mechanism is typically realized in the dual theory by cutting away a region of the worldvolume. This fits remarkably well with the picture that depicts the operators insertions as open D3 brane bubbles: the size of these bubbles is set by $\Delta x_m\sim \mu_{nl}^{-1}$, corresponding to $\xi\sim\xi_0$ in the bulk as dictated by the UV-IR relation~(\ref{eq:uvir}). Hence, the computation leading to~(\ref{eq:pftcorr1}) cuts off two bubbles out of the $x_1$-$x_2$-$x_3$ worldvolume at the location of operator insertions. The modes in the bulk region $\xi<\xi_0$ are entangled with those in the region $\xi>\xi_0$ - mirroring the entanglement of modes insides the D3 brane bubbles with those outside. This is then why we see the vanishing of the rate of convergence of null geodesics at $\xi=\xi_0$. 

In short, the bulk space $\xi>\xi_0$ describes an effective theory - Theory 1 in Figure~\ref{fig:entanglement} - of interacting D3 brane bubbles; the bulk space $\xi<\xi_0$ must then correspond to describing physics {\em inside} the D3 brane bubbles - {\em i.e.} Theory 2 is the effective theory of the internal degrees of freedom of the D3 bubble states. We can offer further evidence for this. Consider cutting the space at $\xi=\xi_2<\xi_0$. The covariant holographic criterion now independently identifies the region $\xi<\xi_2$ as the one projected onto $\xi=\xi_2$. But that translates to an {\em IR cutoff} $\mu_2=\mu_{nl}/\xi_2$. Pushing the cutoff to its limit, $\xi_2=\xi_0$, the region $\xi<\xi_0$ appears like a theory with an IR cutoff $\mu_0$. Indeed, the internal dynamics of the D3 bubbles would necessarily be subject to an IR cutoff, the size of the bubble! If we were to compute correlators in Theory 2, we would use geodesics folding into the region $\xi<\xi_0$, whose length were denoted earlier as $l^{uv}_{\xi_0}$. This length would again involve entanglement information with the modes that are integrated out, now in the region $\xi>\xi_0$; that is, modes outside the D3 bubbles from the dual perspective. 

However, there are no open D3 bubble states in string theory. But there certainly are deformations of the D3 brane worldvolume. The setup makes sense when we realize that the insertions of two non-local operators in the PFT must correspond to the picture cartooned in Figure~\ref{fig:entanglement}(b). The geodesic employed in our computations must be an approximation to a minimal area D3 brane tentacle extending into the bulk. The feet of the tentacle correspond to the operator insertions. The worldvolume theory, with these bubbles subtracted, has dual holographic bulk description given by the $\xi>\xi_0$ region - Theory 1. The internal dynamics of the bubbles is coded into the $\xi<\xi_0$ region - Theory 2. But we now realize that Theory 2 effectively describes D3 brane protrusions! In this particular computation of a 2-point correlator, the protrusion has the shape of a segment $\times S^2$ with the IR cutoff arising from the $S^2$. We can view the unusual topology as one that results from a string folding into the bulk with endpoints on the D3 brane, pulling on the D3 brane worldvolume so as to generate the tentacle in question. We then see a remarkable mechanism to generate non-locality through states that are physically puffed up in size, and a corresponding beautiful realization of the phenomenon from the holographic dual perspective. Theory 1 is an effective description of the PFT local states as well as the interaction dynamics of non-local D3 brane cutouts; while Theory 2 is the effective theory describing the internal dynamics of the D3 protrusions connecting cutouts. The two sectors are not decoupled, and their interaction appears in each of the two theories through quantum entanglement.

There is another twist to this picture. If we focus on correlators of Theory 2 describing the internal degrees of freedom of the D3 tentacles, equation~(\ref{eq:deltaxuv}) and Figure~\ref{fig:dispersionplotcase2} imply that we need to probe distances $\Delta x>\Delta x_m$. The origin of this phenomenon is UV-IR mixing: Theory 2 is the effective theory of the degrees of freedom of the D3 brane tentacle; the corresponding correlators should be viewed as involving insertions at the boundary of the segment. Higher energy corresponds to longer tentacle, hence operator insertions with larger separations! This is reminiscent to the UV-IR mixing that results from considering a large string: higher energy corresponds to longer strings. Put differently, $\Delta x^{ir}$ corresponds to the distance between the insertions in Theory 1, while $\Delta x^{uv}$ is the distance between the insertions {\em along the tentacle} as depicted in Figure~\ref{fig:entanglement}(b), in Theory 2. We note that we always have $\Delta x^{uv}\geq \Delta x^{ir}$ from Figure~\ref{fig:dispersionplotcase2}, with the equality holding at the minimum separation $\Delta x^{uv}=\Delta x^{ir}=\Delta x_m$ as expected.

The causal connection between the two regions of the bulk across the holographic screen - as shown in Figure~\ref{fig:penrosecartan} - suggests that the two theories are not decoupled. The PFT must be the collection of all dynamics, the ones involving puff D3 brane interactions {\em and} the ones involving internal dynamics of the non-local states. This implies that there may be a more complete formulation that does not involve effective dynamics arising from integrating out any degrees of freedom. The entanglement mechanism suggests a theory involving a Fock space with direct product structure. Let us next speculate on the structure of this full theory. 

For states with $Q_4$ R-charge, there is another UV-IR relation at work given by~(\ref{eq:uvir2}). Once the holographic screen is moved into the bulk, the energy read-out for the state is changed: it needs to be measured at the new holographic screen, subject to a different redshift ({\em i.e.} UV-IR) effect. Equation~(\ref{eq:uvir2}) is the energy of a $Q_4$-charged state as measured {\em locally} by an observer located at $\xi=\xi_c$ and $\psi=0$ - with respect to the PFT time coordinate. We rewrite it again here, setting $L_3=0$ for simplicity 
\begin{equation}\label{eq:entanglement}
	E(\mu)=\mu \frac{m\, R}{\sqrt[4]{1+\frac{\mu^6}{\mu_{nl}^6}}}
\end{equation}
For given $E$, there are two solutions for $\mu$; let's call them $E_1$ and $E_2$ (with $\xi_1=\mu_{nl}/E_1$ and $\xi_2=\mu_{nl}/E_2$ in Figure~\ref{fig:energyplotforcase2}).
We propose that the $Q_4$-charged state in the PFT is to be constructed from a product of states in two theories interacting with each other. Let's denote the two states as $|1\rangle$ and $|2\rangle$ with energies $E_1$ and $E_2$ respectively, one in Theory 1 and the other in Theory 2. We construct
\begin{equation}\label{eq:pftstate}
	|E\rangle\equiv |1\rangle\otimes |2\rangle
\end{equation}
$E$ is the energy of the resulting state: fixing $E$, there are two and only two solutions $E_1$ and $E_2$. The interactions between the two theories result in a pairing of states of particular energies to generate an eigenstate of the parent theory.
There is a similar mechanism to this one that arises in non-local theories involving quantum groups. For example, the Moyal plane can be realized by proposing a momentum operator in a new non-local theory that is not diagonalized by the usual direct product of plane waves: instead, a mixed state of the usual momentum eigenstates is employed to construct a new eigenstate of momentum in the non-local theory~\cite{Arzano:2007nx,Arzano:2008yc}. The corresponding correlators exhibit the familiar non-local effects of the non-commutative plane. Our system involves in addition a dynamical element linking up or correlating the two states from which a non-local state is to be build up. In our classical treatment, the correlators involve states localized in space as well as carrying fixed energy. In reality, we should view the position of the operator insertions to be fuzzed up by quantum uncertainty. Hence, we should write the density state matrix of definite position as
\begin{equation}
	|x\rangle\langle x| = \int dE |a(E)|^2 |1\rangle\otimes |2\rangle \langle 1|\otimes \langle 2|
\end{equation}
where the energies of the two states $|1\rangle$ and $|2\rangle$ are the only two solutions to the given function $E(\mu)$ in every term of the integral. The coefficient $a(E)$ determines the profile for the superposition so as to generate a state of definite position at the desired precision (naturally a precision at most of the order of the non-locality scale hidden in the function $E(\mu)$). The correlator of such a state at strong coupling would be
\begin{equation}
	\left|\langle x_1|x_2\rangle\right|	=\sqrt{\mbox{Tr} \left[ |x_1\rangle\langle x_1|x_2\rangle\langle x_2|\right]}\sim e^{-m (l^{ir}_{\xi_0}+l^{uv}_{\xi_0})}\sim e^{-m l^{uv}_{\xi_0}}
\end{equation}
since $l^{uv}_{\xi_0}\gg l^{ir}_{\xi_0}$.
This would be the full result at tree level.
However, integrating out modes of Theory 1 or Theory 2 leads instead to $e^{-m l^{ir}_{\xi_0}}$ or $e^{-m l^{uv}_{\xi_0}}$ respectively, as we discussed earlier. At the leading tree level approximation, integrating out one of the two sectors amounts to just dropping the corresponding exponential factor.

\section{Discussion and Outlook}

PFT apparently involves a remarkably rich regime of string theory, exploring not only non-locality but the foundational principles of the theory as well - the direct interplay between open and closed string sectors. There are several obvious and interesting directions to pursue to develop these ideas further.

\begin{itemize}
	\item Our analysis employed geodesics to delineate the shape of the D3 protrusions. Instead, one could use the DBI action of D3 branes~\cite{Myers:1999ps} to realize the tentacle embedding. One would then compute the area of the tentacle to estimate the 2-point correlator. Such a treatment would improve on our estimates for the correlators, as well as potentially address boundary regimes in the parameter space for $\mathcal{L}_3$ and $\mathcal{P}$ better. We hope to report on this in a future work.
	\item As mentioned in the Introduction, PFT's come in many flavors. A particularly interesting realization may be $2+1$ dimensional PFT. Such a setup would lend itself to more powerful computational tools in determining membrane-like embeddings with interesting topologies. 
	\item While not reported in any detail in this manuscript, we have also looked briefly at the effects of breaking SUSY down to $\mathcal{N}=0$. This is done by turning on additional arbitrary twists in the matrix~(\ref{eq:twist}). Qualitatively, we see no changes in our conclusions. However, such systems are phenomenologically more interesting. As shown in~\cite{Minton:2007fd}, PFT's may be used to compute non-local effects in the Cosmic Microwave Background radiation arising from stringy dynamics in the primordial plasma. Less SUSY, or even the possibility of a dynamical cascade through various SUSY realizations of the PFT, can play an important role in this program.
	\item The primary question still remains unanswered: what is a full description of the PFT? one that does not involve entanglement and effective theories, but the full mess. Does the PFT have a complete description with a product structure for its states? Is there a role played by quantum groups in this description? These are hard questions with deep implications to open-closed string duality.
	\item Another interesting direction involves the general idea of using entanglement to realize non-locality in a theory. This is in the spirit of black hole entanglement phenomena\cite{Hawking:2000da,Maldacena:2001kr,Brustein:2005vx}, where a hidden sector behind the horizon plays a role in explaining non-local effects in horizon dynamics. Here, we see a possibly more mundane realization of this idea in a non-gravitational field theory. It would be interesting if one can cook up toy models where two theories are combined dynamically in a direct product structure, and entanglement is made to seed non-local effects at low energies.
\end{itemize}

We hope to visit some of these topics in the near future.

\section{Acknowledgments}
V.S. thanks the Caltech theory group for hospitality. This work was supported by Research Corporation grant No. CC6483 and the Baker Foundation. 

\newpage {\Large \bf Appendices}

\vspace{0.5in} {\Large \bf Appendix A: Some details about the R-symmetry} \vspace{0.25in}

To put the various R-charges explored in the main text in the context of the coordinates of the bulk spacetime, we list in this appendix the matrix transformations that twist the $\chi$-$\theta$-$\phi$-$\varphi$ angles. We use a representation of the matrices in the cartesian basis mentioned in Section 1 - with coordinates labeled as $y_{1}\cdots y_6$ - related to the polar coordinates $y_{1,2}\rightarrow r_1,\phi_1$, $y_{3,4}\rightarrow r_2,\phi_2$, and $y_{5,6}\rightarrow r_3,\phi_3$. The angles appearing in the metric are then related to these coordinates by $\phi=\phi_1$, $\varphi=\phi_1-\phi_2$, $\tan \chi=y_5/y_6$, $\sin \theta=2\,r_1 r_2/(r_1^2+r_2^2)$, $\tan \psi=\sqrt{r_1^2+r_2^2}/\sqrt{r_1^2+r_2^2+r_3^2}$. With these conventions, the matrices become:
\begin{equation}
	{\bf Q_0} = \left( 
	\begin{array}{cccccc}
		0 & 0 & 0 & 0 & 0 & 0 \\
		0 & 0 & 0 & 0 & 0 & 0 \\
		0 & 0 & 0 & 0 & 0 & 0 \\
		0 & 0 & 0 & 0 & 0 & 0 \\
		0 & 0 & 0 & 0 & 0 & 1 \\
		0 & 0 & 0 & 0 & -1 & 0 
	\end{array}
	\right) 
\end{equation}
\begin{eqnarray}
	&{\bf Q_1} = \frac{1}{2}\left( 
	\begin{array}{cccccc}
		0 & 0 & 0 & -1 & 0 & 0 \\
		0 & 0 & -1 & 0 & 0 & 0 \\
		0 & 1 & 0 & 0 & 0 & 0 \\
		1 & 0 & 0 & 0 & 0 & 0 \\
		0 & 0 & 0 & 0 & 0 & 0 \\
		0 & 0 & 0 & 0 & 0 & 0 
	\end{array}
	\right) \ &{\bf Q_2} = \frac{1}{2}\left( 
	\begin{array}{cccccc}
		0 & -1 & 0 & 0 & 0 & 0 \\
		1 & 0 & 0 & 0 & 0 & 0 \\
		0 & 0 & 0 & -1 & 0 & 0 \\
		0 & 0 & 1 & 0 & 0 & 0 \\
		0 & 0 & 0 & 0 & 0 & 0 \\
		0 & 0 & 0 & 0 & 0 & 0 
	\end{array}
	\right) \nonumber \\
	&{\bf Q_3} = \frac{1}{2}\left( 
	\begin{array}{cccccc}
		0 & 0 & -1 & 0 & 0 & 0 \\
		0 & 0 & 0 & 1 & 0 & 0 \\
		1 & 0 & 0 & 0 & 0 & 0 \\
		0 & -1 & 0 & 0 & 0 & 0 \\
		0 & 0 & 0 & 0 & 0 & 0 \\
		0 & 0 & 0 & 0 & 0 & 0 
	\end{array}
	\right)\ &{\bf Q_4} = \frac{1}{2}\left( 
	\begin{array}{cccccc}
		0 & -1 & 0 & 0 & 0 & 0 \\
		1 & 0 & 0 & 0 & 0 & 0 \\
		0 & 0 & 0 & 1 & 0 & 0 \\
		0 & 0 & -1 & 0 & 0 & 0 \\
		0 & 0 & 0 & 0 & 0 & 0 \\
		0 & 0 & 0 & 0 & 0 & 0 
	\end{array}
	\right) \nonumber 
\end{eqnarray}
Hence, ${\bf Q_0}$ translates only $\chi$, and ${\bf Q_4}$ translates $\varphi$ twice as much as it does $\phi$.

\bibliographystyle{utphys}
% \bibliography{biblio}

\begin{thebibliography}{10}

\bibitem{Seiberg:1999vs}
N.~Seiberg and E.~Witten, ``{String theory and noncommutative geometry},'' {\em
  JHEP} {\bf 09} (1999) 032, \href{http://xxx.lanl.gov/abs/hep-th/9908142}{{\tt
  hep-th/9908142}}.

\bibitem{Hashimoto:1999ut}
A.~Hashimoto and N.~Itzhaki, ``Non-commutative Yang-Mills and the AdS/CFT
  correspondence,'' {\em Phys. Lett.} {\bf B465} (1999) 142--147,
  \href{http://xxx.lanl.gov/abs/hep-th/9907166}{{\tt hep-th/9907166}}.

\bibitem{Hashimoto:1999yj}
A.~Hashimoto and N.~Itzhaki, ``On the hierarchy between non-commutative and
  ordinary supersymmetric Yang-Mills,'' {\em JHEP} {\bf 12} (1999) 007,
  \href{http://xxx.lanl.gov/abs/hep-th/9911057}{{\tt hep-th/9911057}}.

\bibitem{Cai:1999aw}
R.-G. Cai and N.~Ohta, ``{On the thermodynamics of large N non-commutative
  super Yang-Mills theory},'' {\em Phys. Rev.} {\bf D61} (2000) 124012,
  \href{http://xxx.lanl.gov/abs/hep-th/9910092}{{\tt hep-th/9910092}}.

\bibitem{Maldacena:1999mh}
J.~M. Maldacena and J.~G. Russo, ``Large N limit of non-commutative gauge
  theories,'' {\em JHEP} {\bf 09} (1999) 025,
  \href{http://xxx.lanl.gov/abs/hep-th/9908134}{{\tt hep-th/9908134}}.

\bibitem{Seiberg:2000ms}
N.~Seiberg, L.~Susskind, and N.~Toumbas, ``Strings in background electric
  field, space/time noncommutativity and a new noncritical string theory,''
  {\em JHEP} {\bf 06} (2000) 021,
  \href{http://xxx.lanl.gov/abs/hep-th/0005040}{{\tt hep-th/0005040}}.

\bibitem{Klebanov:2000pp}
I.~R. Klebanov and J.~M. Maldacena, ``1+1 dimensional NCOS and its U(N) gauge
  theory dual,'' {\em Int. J. Mod. Phys.} {\bf A16} (2001) 922--935,
  \href{http://xxx.lanl.gov/abs/hep-th/0006085}{{\tt hep-th/0006085}}.

\bibitem{Sahakian:2001xz}
V.~Sahakian, ``The large M limit of non-commutative open strings at strong
  coupling,'' {\em Nucl. Phys.} {\bf B621} (2002) 62--100,
  \href{http://xxx.lanl.gov/abs/hep-th/0107180}{{\tt hep-th/0107180}}.

\bibitem{Harmark:2000wv}
T.~Harmark, ``Supergravity and space-time non-commutative open string theory,''
  {\em JHEP} {\bf 07} (2000) 043,
  \href{http://xxx.lanl.gov/abs/hep-th/0006023}{{\tt hep-th/0006023}}.

\bibitem{Gopakumar:2000ep}
R.~Gopakumar, S.~Minwalla, N.~Seiberg, and A.~Strominger, ``{OM Theory in
  Diverse Dimensions},'' {\em JHEP} {\bf 08} (2000) 008,
  \href{http://xxx.lanl.gov/abs/hep-th/0006062}{{\tt hep-th/0006062}}.

\bibitem{Bergman:2000cw}
A.~Bergman and O.~J. Ganor, ``Dipoles, twists and noncommutative gauge
  theory,'' {\em JHEP} {\bf 10} (2000) 018,
  \href{http://xxx.lanl.gov/abs/hep-th/0008030}{{\tt hep-th/0008030}}.

\bibitem{Bergman:2001rw}
A.~Bergman, K.~Dasgupta, O.~J. Ganor, J.~L. Karczmarek, and G.~Rajesh,
  ``{Nonlocal field theories and their gravity duals},'' {\em Phys. Rev.} {\bf
  D65} (2002) 066005, \href{http://xxx.lanl.gov/abs/hep-th/0103090}{{\tt
  hep-th/0103090}}.

\bibitem{Dasgupta:2001zu}
K.~Dasgupta and M.~M. Sheikh-Jabbari, ``{Noncommutative dipole field
  theories},'' {\em JHEP} {\bf 02} (2002) 002,
  \href{http://xxx.lanl.gov/abs/hep-th/0112064}{{\tt hep-th/0112064}}.

\bibitem{Alishahiha:2002ex}
M.~Alishahiha and H.~Yavartanoo, ``{Supergravity description of the large N
  noncommutative dipole field theories},'' {\em JHEP} {\bf 04} (2002) 031,
  \href{http://xxx.lanl.gov/abs/hep-th/0202131}{{\tt hep-th/0202131}}.

\bibitem{Ganor:2006ub}
O.~J. Ganor, ``{A New Lorentz Violating Nonlocal Field Theory From String-
  Theory},'' {\em Phys. Rev.} {\bf D75} (2007) 025002,
  \href{http://xxx.lanl.gov/abs/hep-th/0609107}{{\tt hep-th/0609107}}.

\bibitem{Ganor:2007qh}
O.~J. Ganor, A.~Hashimoto, S.~Jue, B.~S. Kim, and A.~Ndirango, ``{Aspects of
  puff field theory},'' {\em JHEP} {\bf 08} (2007) 035,
  \href{http://xxx.lanl.gov/abs/hep-th/0702030}{{\tt hep-th/0702030}}.

\bibitem{Haque:2008vm}
S.~S. Haque and A.~Hashimoto, ``{Microscopic Formulation of Puff Field
  Theory},'' {\em JHEP} {\bf 05} (2008) 040,
  \href{http://xxx.lanl.gov/abs/0801.4354}{{\tt 0801.4354}}.

\bibitem{Minton:2007fd}
G.~Minton and V.~Sahakian, ``{A new mechanism for non-locality from string
  theory: UV-IR quantum entanglement and its imprints on the CMB},'' {\em Phys.
  Rev.} {\bf D77} (2008) 026008, \href{http://xxx.lanl.gov/abs/0707.3786}{{\tt
  0707.3786}}.

\bibitem{Maldacena:1997re}
J.~M. Maldacena, ``The large N limit of superconformal field theories and
  supergravity,'' {\em Adv. Theor. Math. Phys.} {\bf 2} (1998) 231--252,
  \href{http://xxx.lanl.gov/abs/hep-th/9711200}{{\tt hep-th/9711200}}.

\bibitem{Witten:1998qj}
E.~Witten, ``Anti-de Sitter space and holography,'' {\em Adv. Theor. Math.
  Phys.} {\bf 2} (1998) 253--291,
  \href{http://xxx.lanl.gov/abs/hep-th/9802150}{{\tt hep-th/9802150}}.

\bibitem{Gubser:1998bc}
S.~S. Gubser, I.~R. Klebanov, and A.~M. Polyakov, ``Gauge theory correlators
  from non-critical string theory,'' {\em Phys. Lett.} {\bf B428} (1998)
  105--114, \href{http://xxx.lanl.gov/abs/hep-th/9802109}{{\tt
  hep-th/9802109}}.

\bibitem{Cai:2006tda}
R.~G.~Cai and N.~Ohta, ``Holography and D3-branes in Melvin universes,'' {\em Phys. Rev.} {\bf D73} (2006)
  106009, \href{http://xxx.lanl.gov/abs/hep-th/0601044}{{\tt
  hep-th/0601044}}.

\bibitem{Hashimoto:2005hy}
Hashimoto, Akikazu and Thomas, Keith, ``Non-commutative gauge theory on D-branes in Melvin universes,'' {\em JHEP} {\bf 01} (2006) 083, \href{http://xxx.lanl.gov/abs/hep-th/0511197}{{\tt
  hep-th/0511197}}.

\bibitem{Dhokarh:2007ry}
Dhokarh, Danny and Hashimoto, Akikazu and Haque, Sheikh
                  Shajidul, ``Non-commutativity and Open Strings Dynamics in Melvin Universes,'' {\em JHEP} {\bf 08} (2007) 027, \href{http://xxx.lanl.gov/abs/0704.1124}{{\tt 0704.1124}}.

\bibitem{Dhokarh:2008ki}
Dhokarh, Danny and Haque, Sheikh Shajidul and Hashimoto, Akikazu, ``Melvin Twists of global AdS5xS5 and their Non-Commutative Field Theory Dual,'' {\em JHEP} {\bf 08} (2008) 084, \href{http://xxx.lanl.gov/abs/0801.3812}{{\tt 0801.3812}}.

\bibitem{Peet:1998wn}
A.~W. Peet and J.~Polchinski, ``UV/IR relations in AdS dynamics,'' {\em Phys.
  Rev.} {\bf D59} (1999) 065011,
  \href{http://xxx.lanl.gov/abs/hep-th/9809022}{{\tt hep-th/9809022}}.

\bibitem{Bousso:1999xy}
R.~Bousso, ``A covariant entropy conjecture,'' {\em JHEP} {\bf 07} (1999) 004,
  \href{http://xxx.lanl.gov/abs/hep-th/9905177}{{\tt hep-th/9905177}}.

\bibitem{Ryu:2006bv}
S.~Ryu and T.~Takayanagi, ``Holographic derivation of entanglement entropy from
  AdS/CFT,'' {\em Phys. Rev. Lett.} {\bf 96} (2006) 181602,
  \href{http://xxx.lanl.gov/abs/hep-th/0603001}{{\tt hep-th/0603001}}.

\bibitem{Ryu:2006ef}
S.~Ryu and T.~Takayanagi, ``{Aspects of holographic entanglement entropy},''
  {\em JHEP} {\bf 08} (2006) 045,
  \href{http://xxx.lanl.gov/abs/hep-th/0605073}{{\tt hep-th/0605073}}.

\bibitem{Hubeny:2007xt}
V.~E. Hubeny, M.~Rangamani, and T.~Takayanagi, ``A covariant holographic
  entanglement entropy proposal,''
  \href{http://xxx.lanl.gov/abs/arXiv:0705.0016 [hep-th]}{{\tt arXiv:0705.0016
  [hep-th]}}.

\bibitem{Nishioka:2009un}
T.~Nishioka, S.~Ryu, and T.~Takayanagi, ``{Holographic Entanglement Entropy: An
  Overview},'' \href{http://xxx.lanl.gov/abs/0905.0932}{{\tt 0905.0932}}.

\bibitem{VanRaamsdonk:2009ar}
M.~Van~Raamsdonk, ``{Comments on quantum gravity and entanglement},''
  \href{http://xxx.lanl.gov/abs/0907.2939}{{\tt 0907.2939}}.

\bibitem{Arzano:2007nx}
M.~Arzano, ``{Quantum fields, non-locality and quantum group symmetries},''
  {\em Phys. Rev.} {\bf D77} (2008) 025013,
  \href{http://xxx.lanl.gov/abs/0710.1083}{{\tt 0710.1083}}.

\bibitem{Arzano:2008yc}
M.~Arzano, A.~Hamma, and S.~Severini, ``{Hidden entanglement at the Planck
  scale: loss of unitarity and the information paradox},''
  \href{http://xxx.lanl.gov/abs/0806.2145}{{\tt 0806.2145}}.

\bibitem{Balasubramanian:1998sn}
V.~Balasubramanian, P.~Kraus, and A.~E. Lawrence, ``Bulk vs. boundary dynamics
  in anti-de sitter spacetime,'' {\em Phys. Rev.} {\bf D59} (1999) 046003,
  \href{http://xxx.lanl.gov/abs/hep-th/9805171}{{\tt hep-th/9805171}}.

\bibitem{Balasubramanian:1998de}
V.~Balasubramanian, P.~Kraus, A.~E. Lawrence, and S.~P. Trivedi, ``Holographic
  probes of anti-de sitter space-times,'' {\em Phys. Rev.} {\bf D59} (1999)
  104021, \href{http://xxx.lanl.gov/abs/hep-th/9808017}{{\tt hep-th/9808017}}.

\bibitem{Rozali:2000np}
M.~Rozali and M.~Van~Raamsdonk, ``Gauge invariant correlators in
  non-commutative gauge theory,'' {\em Nucl. Phys.} {\bf B608} (2001) 103--124,
  \href{http://xxx.lanl.gov/abs/hep-th/0012065}{{\tt hep-th/0012065}}.

\bibitem{Gross:2000ba}
D.~J. Gross, A.~Hashimoto, and N.~Itzhaki, ``Observables of non-commutative
  gauge theories,'' {\em Adv. Theor. Math. Phys.} {\bf 4} (2000) 893--928,
  \href{http://xxx.lanl.gov/abs/hep-th/0008075}{{\tt hep-th/0008075}}.

\bibitem{Myers:1999ps}
Myers,~Robert C., ``Dielectric-branes,'' {\em JHEP} {\bf 12} (1999) 022,
  \href{http://xxx.lanl.gov/abs/hep-th/9910053}{{\tt hep-th/9910053}}.

\bibitem{Hawking:2000da}
S.~Hawking, J.~M. Maldacena, and A.~Strominger, ``Desitter entropy, quantum
  entanglement and ads/cft,'' {\em JHEP} {\bf 05} (2001) 001,
  \href{http://xxx.lanl.gov/abs/hep-th/0002145}{{\tt hep-th/0002145}}.

\bibitem{Maldacena:2001kr}
J.~M. Maldacena, ``Eternal black holes in anti-de-sitter,'' {\em JHEP} {\bf 04}
  (2003) 021, \href{http://xxx.lanl.gov/abs/hep-th/0106112}{{\tt
  hep-th/0106112}}.

\bibitem{Brustein:2005vx}
R.~Brustein, M.~B. Einhorn, and A.~Yarom, ``Entanglement interpretation of
  black hole entropy in string theory,'' {\em JHEP} {\bf 01} (2006) 098,
  \href{http://xxx.lanl.gov/abs/hep-th/0508217}{{\tt hep-th/0508217}}.

\end{thebibliography}

\providecommand{\href}[2]{#2}\begingroup\raggedright\endgroup

\end{document}